\documentclass[journal]{IEEEtran}
\IEEEoverridecommandlockouts

\usepackage{amsmath,amssymb,amsfonts,bm}
\usepackage{cite}
\usepackage{color}
\usepackage{subfigure}
\usepackage{epstopdf}
\usepackage{booktabs}
\usepackage{rotating}
\usepackage{stfloats}
\usepackage[breaklinks,colorlinks,linkcolor=blue,citecolor=blue, anchorcolor=blue]{hyperref}

\allowdisplaybreaks[4]

\begin{document}

\title{On the Mutual Interference between Spaceborne SARs: Modeling, Characterization, and Mitigation}

\author{Huizhang~Yang,
    Mingliang~Tao,
	Shengyao~Chen,
    Feng~Xi,
	and Zhong~Liu
	\thanks{
    H. Yang, S. Chen, F. Xi, and Z. Liu are with the School of Electronic and Optical Engineering, Nanjing University of Science and Technology, Nanjing 210094, China.
    (email: hzyang@njust.edu.cn or huizhang.yang@foxmail.com; chenshengyao@njust.edu.cn; xifeng@njust.edu.cn;eezliu@njust.edu.cn)

    M. Tao is with the School of Electronics and Information, Northwestern Polytechnical University, Xi'an 710072, China. (email: mltao@foxmail.com)
    }
}
\maketitle
\begin{abstract}
As the radio spectrum available to spaceborne synthetic aperture radar (SAR) is restricted to certain limited frequency intervals, there are many different spaceborne SAR systems sharing common frequency bands. Due to this reason, it is reported that two spaceborne SARs at orbit cross positions can potentially cause severe mutual interference. Specifically, the transmitting signal of a SAR, typically linear frequency modulated (LFM), can be directly received by the side or back lobes of another SAR's antenna, causing radiometric artefacts in the focused image. This paper tries to model and characterize the artefacts, and study efficient methods for mitigating them. To this end, we formulate an analytical model for describing the artefact, which reveals that the mutual interference can introduce a two-dimensional LFM radiometric artefact in image domain with a limited spatial extent. We show that the artefact is low-rank based on range-azimuth decoupling analysis and two-dimensional high-order Taylor expansion. Based on the low rank model, we show that two methods, i.e., principal component analysis and its robust variant, can be adopted to efficiently mitigate the artefact via processing in image domain. The former method has the advantage of fast processing speed, for example, a sub-swath of Sentinel-1 interferometric wide swath image can be processed within $70$ seconds via block-wise processing, whereas the latter provides improved accuracy for sparse point-like scatterers. Experiment results demonstrate that the radiometric artefacts caused by  mutual interference in Sentinel-1 level-1 images can be efficiently mitigated via the proposed methods.

\end{abstract}

\IEEEpeerreviewmaketitle

\section{Introduction}

\begin{figure}
  \centering
  \subfigure{\includegraphics[width=8.8cm]{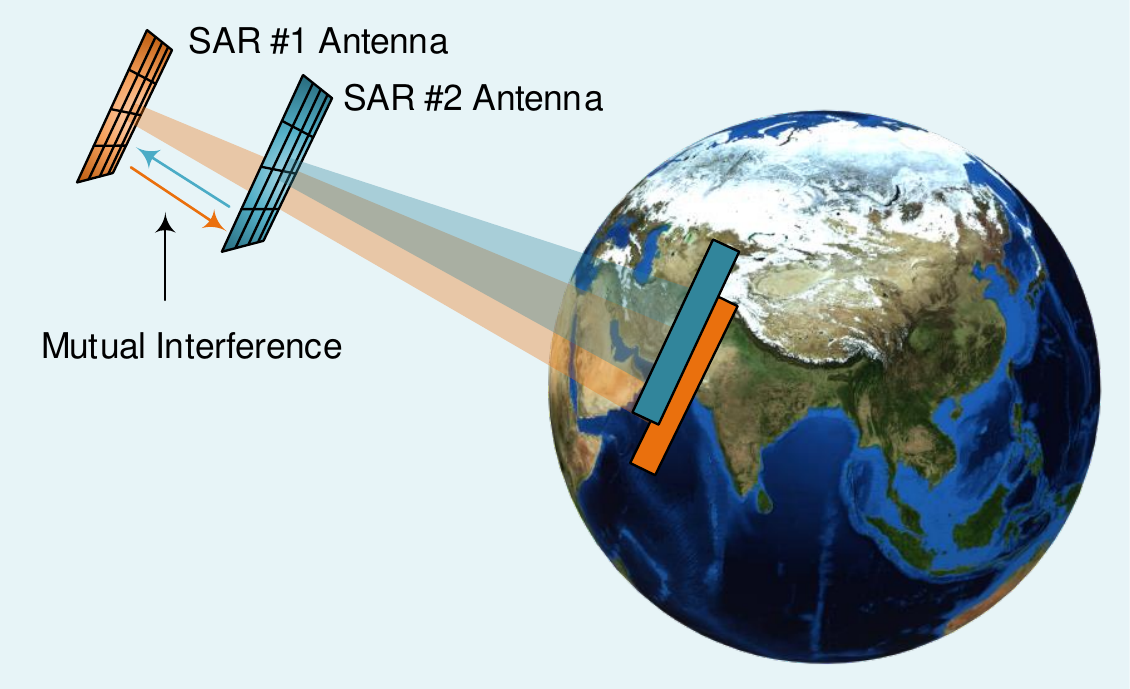}}
  \caption{Illustration of spaceborne SAR geometry with mutual interference. The radar signal transmitted by a SAR is directly received by another SAR's antenna, causing linear-frequency-modulated interference. Modified from \cite{Mingliang2019Mitigation}.}\label{jam}
\end{figure}

\begin{figure}
  \centering
\subfigure{\includegraphics[width=8.8cm]{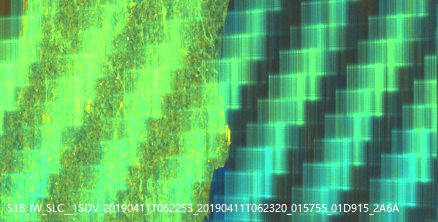}}
\subfigure{\includegraphics[width=8.8cm]{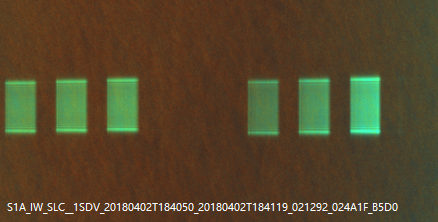}}
  \caption{Examples of radiometric artefacts reported to be caused by spaceborne SARs. Extracted from S-1 IW quick-look images. The image vertical (azimuth) span is about one S-1 IW burst length.}\label{ex}
\end{figure}

Mounted on a satellite, a spaceborne synthetic aperture radar (SAR) maps the Earth surface via actively transmitting radio signals. Due to its important civil and military applications, many institutes as well as business companies have entered this fields with a number of spaceborne SARs been launched and in operation. According to the International Telecommunications Union (ITU), the frequency allocation for spaceborne active sensor is restricted within certain limited radio frequency (RF) intervals. Therefore, there are many different SAR systems operating with the same RF bands. For example, Sentinel-1A/B (S-1), Radarsat-2, and Radarsat Constellation Mission-1/2/3 are currently operating at a central frequency of 5405 MHz with an RF bandwidth of 100 MHz. Furthermore, Gaofen-3 is operating at 5400 MHz with an RF bandwidth 240 MHz, overlapping the same frequency range as the aforementioned missions. As a consequence, the transmitted linear-frequency-modulated (LFM) signal from one SAR might be directly received by the other SAR's antenna (and vice versa) at orbit cross positions, causing mutual interference, as illustrated in Fig. \ref{jam}.
The unwanted interference can introduce severe radiometric artefacts in the focused images, as reported by the S-1 Mission Performance Center (MPC) \cite{s1mpc,s1mpc2}. Fig. \ref{ex} shows some examples of these radiometric artefacts observed in S-1 interferometric wide swath (IW) images possibly due to spaceborne SARs or surface RFI sources. In the images, it is shown that the radiometric artefacts have high intensity, burying the underlying scene reflectivity. This problem is adverse to subsequent high-level processing or image interpretation applications, like interferometry, polarimetry, target detection and classification. In the future, the mutual interference will only get worse with the advent of new missions, increasing use of dusk/dawn orbits for SAR satellites, and small companies doing microwave remote sensing. Therefore, it is important to model, characterize, and then mitigate the mutual interference.

In the literature, RF interference (RFI) mitigation for SAR is an active research topic and has been widely investigated. Reported works on this topic can be mainly categorized by the RFI bandwidth into two classes. One is designed for narrowband RFI and the other is for the wideband case.
For mitigating narrowband RFI, a simple and widely used method is the range spectrum notch filtering \cite{Koutsoudis1995rf,Lord1999Efficient,Meyer2013}, of which the performance is usually limited. Advanced techniques, like  subspace-based methods, can be applied to achieve better performance by means of extracting and then removing the RFI\cite{Zhou2007,Zhou2013na,Tao2014}. In recent years, it is shown that the RFI mitigation performance can be further improved via optimization methods like sparse recovery\cite{Nguyen2016Efficient,8515096}, sparse frequency estimation\cite{ren2018rfi}, and low-rank matrix recovery\cite{Nguyen2012Robust,JoyRob2014,Nguyen2015Radio,SuNarrow2017,Yang2020SAR,9127088,Yang2020}. These state-of-art methods impose the narrowband assumption on interference, which excludes the case of wideband RFI, like LFM interference.

For wideband RFI mitigation, a lot of efforts have been made by many researchers\cite{Nguyen2004,Tao2016,huang2020,8282759}. Nguyen et al. proposed an adaptive coherent method for mitigating multiple wideband RFI sources\cite{Nguyen2004}. Zhang et al. designed a time-frequency filtering approach for dealing with both narrowband and wideband RFI that have sharp peaks in  time-frequency domain \cite{zhang2011}. Tao et al. performed systematic analysis for wideband RFI modeling, adaptive detection, and mitigation via time-frequency and eigen-subspace filtering \cite{Tao2016}. In\cite{Tao2016}, the wideband RFI detection was firstly performed on a pulse-by-pulse basis using a statistical criteria under the assumption that the radar echo is complex Gaussian, and then filtering was applied to the pulses that contain RFI. For sparse point scatterers, the assumption is not valid and the two filtering approaches may fail. As one can imagine, the radar echo of a single strong point scatterer is also an LFM signal, and thus it is hard for a detector to distinguish between the useful LFM echo and the unwanted LFM interference.

It is common to all the above work that the interference mitigation is performed in raw echo domain rather than in image domain. A reason for this fact is that theoretical characterization of interference signatures in image domain is difficult, mainly due to the complicated process of SAR image focusing. If the raw data is unavailable and there is only a focused image, some post processing can be applied to transform the image into appropriate domains followed by filtering. For example, Reigber et al. used inverse focusing processing to transform focused images into range-frequency azimuth-time domain and then traditional notch filtering was applied to cut off narrowband RFI pulse-by-pulse\cite{Reigber2005}. This idea can be extended to deal with wideband RFI by incorporating with time-varying filtering, but it requires prior knowledge about some parameters involved in the focusing process, which may be not known by the end user. Even if these parameters are available, this method can be very complicated in some advanced imaging modes like terrain observation by progressive scan (TOPS), because focusing in these modes is more tedious than the standard stripmap processing.

Motivated by these issues, in this paper we first try to derive an analytical model for characterizing spaceborne SAR mutual interference signatures in image domain, as we observed in the literature that there lack interference modeling and characterization in this domain. We demonstrate that, under mild approximations, a single LFM pulse transmitted by a SAR and received by another SAR as  interference will introduce a two-dimensional LFM artefact with a limited range-azimuth extent in image domain. This result is also valid when the RFI is from other sources (e.g., military air defense radars) as long as it is LFM. The analytical model explains why the LFM interference observed in SAR images admits certain structured pattern, as the examples in Fig. \ref{ex}.

Second, we show that the LFM interference artefact is low-rank based on range-azimuth decoupling analysis and two-dimensional high-order Taylor expansion. We test the proposed low rank modeling using  numerical simulations in a C-band case, and results demonstrate that the proposed low rank model is efficient.

Third, using the proposed low rank model, we show that two simple yet efficient techniques can be applied to remove LFM interference artefacts in image domain. The first one is the principal component analysis (PCA) for the case where the underlying reflectivity can be viewed as complex Gaussian, and the second one is the robust PCA (RPCA)\cite{candes2011robust}, for the situation where the underlying reflectivity has strong point scatterers. The former has an advantage of computational efficiency, whereas the latter generally provides better performance, since practical scene reflectivity often deviates from Gaussian distribution. The two methods only need to be applied to the image patches that contain interference artefacts, which can be easily identified by human eye. In addition, they can be implemented in a block-wise manner to boost processing speed. These benefits are important from the computational perspective since a whole SAR image usually has a quite large size. We test the two methods on  S-1 single look complex (SLC) images which are severely corrupted by LFM interference. Results show that the interference-induced artefacts can be significantly mitigated in image domain based on the proposed low rank model.

The remainder of this paper is organized as follows. Section II introduces SAR signal model and some preliminary knowledge. Section III presents theoretical model for the image-domain artefact caused by LFM interference. Section IV discusses  the spatial and spectral characteristics of the interference artefact. Section V proposes the low rank model for the interference artefact. Section VI introduces PCA and RPCA for interference artefact mitigation. Section VII provides real-data examples with S-1 data. Section VIII concludes this paper.

\section{Preliminaries}

In this section we briefly review the basic SAR signal model, and introduce the principle of stationary phase and an approximated version of the omega-K algorithm as preliminary knowledge for subsequent modeling in the next section.

\subsection{SAR Signal Model}

SAR operates by sequentially transmitting and receiving  radio signals to collect echoes backscattered from the region of interest illuminated by the radar beam. The time within a pulse observation interval is called range time, and that across different pulse observation interval  is referred to as azimuth time. Denoting the range and azimuth times by $\tau$ and $\eta$, respectively, we can express the radar echo of a point scatterer as
\begin{equation}\label{}
 s(\eta,\tau)= \gamma_0 h\left(\tau-\frac{2R(\eta)}{c}\right)
 \exp\left(j2\pi f_0\left(\tau-\frac{2R(\eta)}{c}\right)\right),
\end{equation}
where $\gamma_0$ is the complex amplitude of the scatterer's echo, $h(\tau)$ is the transmitted radar signal, $f_0$ is the carrier frequency of the transmitted radar signal, $R(\eta)$ is its slant range history and $c$ is the speed of light. Defining the zero-Doppler azimuth position and minimum slant range of the scatterer as $x_0$ and $R_0$, respectively, we can express $R(\eta)$ as
$R(\eta)=\sqrt{(V\eta-x_0)^2+R_0^2}$,
where $V$ is the effective radar velocity.

For spaceborne SARs, the transmitted signal is usually an LFM pulse, which is defined using a pulse duration $T$ and linear frequency modulation (FM) rate $K_{\rm r}$ as
\begin{equation}\label{}
  h(\tau)=\operatorname{rect}\left(\frac{\tau}{T}\right)\exp\left(j\pi K_{\rm r}\tau^2\right),
\end{equation}
where $\operatorname{rect}(\cdot)$ is a rectangular function. For simplicity, we have ignored the constant term that represents the pulse energy in the LFM signal. The mutual interference caused by the transmitted signal from another SAR also admits the above signal form but with a generally different pulse duration and FM rate. To distinguish between the actively transmitted LFM pulse and the passively received interference, we use a subscript `i' to mark the interference LFM pulse as
\begin{equation}\label{}
  h_{\rm i}(\tau)=\operatorname{rect}\left(\frac{\tau}{T_{\rm i}}\right)\exp\left(j\pi K_{\rm i}\tau^2\right).
\end{equation}

\subsection{Principle of Stationary Phase}

In the following we will frequently use the principle of stationary phase (POSP), which is a fundamental tool for analyzing SAR signals. It states that the major contribution to the energy of an integral, which is assumed to have rapidly varying phase but slowly varying envelope, comes from  stationary phase points. According to this principle, the Fourier transform of a signal $g(t)=w(t)\exp(j\phi(t))$ can be easily derived if the real-valued envelope $w(t)$ is a slowly varying function of $t$. Specifically, for the Fourier transform
\begin{equation}\label{}
  G(f)=\int w(t)\exp\left(j\phi(t)-j2\pi f t)\right)dt,
\end{equation}
we define the phase in the integral as $\theta(t)=\phi(t)-2\pi ft$, and then take the zero derivative of $\theta(t)$ to obtain the stationary phase point, i.e.,
\begin{equation}\label{}
\frac{d\phi(t)}{dt}-2\pi f=0.
\end{equation}
 Then inverting this equation to get $t=u(f)$ as a function of $f$, and inserting $t=u(f)$ into the integral, we can obtain the desired result
  \begin{equation}\label{}
    G(f)=Cw\big(u(f)\big)\exp\big(j\phi(u(f))-j2\pi fu(f)\big),
  \end{equation}
where $C$ is a constant term. For simplicity, we will ignore $C$ in the following.

 According to POSP, the Fourier transform of the LFM interference is given as follows
 \begin{equation}\label{}
   H_{\rm i}(f_{\tau})=\operatorname{rect}\left(\frac{f_{\tau}}{K_{\rm i}T_{\rm i}}\right)\exp\left(-j\frac{\pi f_{\tau}^2}{K_{\rm i}}\right).
 \end{equation}

\subsection{Omega-K Algorithm}

Omega-K is a wavenumber-domain focusing algorithm for SAR. It has an accurate version as well as an approximated form. In the following we will use the approximated form to derive a closed-form expression for the image-domain interference artefact. For the ease of discussion, we take the zero-Doppler time of a point scatterer with minimum slant range $R_0$ as the reference, i.e., its position in azimuth-slant range coordinate is $(0,R_0)$. The wavenumber-domain expression of the scatterer's baseband impulse response is
\begin{equation}\label{}
  S_{\rm 2df}(f_{\eta},f_{\tau})=A W_{\eta}(f_{\eta}-f_{\eta_{\rm c}})W_{\tau}(f_{\tau})\exp\left(j\theta_{\rm 2df}\right),
\end{equation}
where $A$ is a constant, $f_{\eta_{\rm c}}$ is the Doppler centroid, $W_{\eta}(\cdot)$ and $W_{\tau}(\cdot)$ are the range and azimuth frequency envelope, respectively, and $\theta_{\rm 2df}$ is the phase of $S_{\rm 2df}(f_{\eta},f_{\tau})$.
Ignoring high-order range-azimuth coupling, we can express $\theta_{\rm 2df}$  as
\begin{equation}\label{}
\theta_{\rm 2df}=-\frac{4\pi R_0}{c}\sqrt{(f_0+f_{\tau})^2-\frac{c^2f_{\eta}^2}{4V^2}}-\frac{\pi f_{\tau}^2}{K_{\rm r}},
\end{equation}
where $D(f_{\eta},V)$ is a hyperbolic range cell mitigation parameter defined as
\begin{equation}\label{dfv}
  D(f_{\eta},V)=\sqrt{1-\frac{c^2f_{\eta}^2}{4V^2f_0^2}}.
\end{equation}
Based on the above expression, the approximated version of omega-K focuses SAR echo data via five steps:
\begin{enumerate}
  \item Transform echo data into wavenumber domain via two-dimensional Fourier transform.
  \item Multiply the wavenumber-domain spectrum by a reference function $\exp\left(j\theta_{\rm ref}\right)$ where
\begin{equation}\label{}
  \theta_{\rm ref}=\frac{4\pi R_{\rm ref}}{c}\sqrt{(f_0+f_{\tau})^2-\frac{c^2f_{\eta}^2}{4V^2}}+\frac{\pi f_{\tau}^2}{K_{\rm r}}.
\end{equation}
  \item Apply range inverse Fourier transform to obtain range-focused data in the range-Doppler domain.
  \item Multiply the range-Doppler spectrum at each range gate $R_0$ by azimuth matched filtering function $\exp(j\theta_{\rm mfa})$ where
    \begin{equation}\label{}
  \theta_{\rm mfa}=\frac{4\pi(R_0-R_{\rm ref})}{c}f_0D(f_{\eta},V),
\end{equation}
and $R_{\rm ref}$ is the reference slant range usually chosen using the scene center position.
  \item Apply azimuth inverse Fourier transform to accomplish azimuth focusing.
\end{enumerate}

\section{Analytical Model for Mutual Interference Artefact}

Based on the above preliminary knowledge, in this section we will derive an analytical model for characterizing the image-domain artefact caused by an LFM interference, as presented in the following.

\subsection{LFM Mutual Interference Observed in Two-Dimensional SAR Echo}
To facilitate our analysis, we take the azimuth time of an LFM interference as the  azimuth time origin $\eta=0$. The LFM interference is a scaled and delayed version of $h_{\rm i}(\tau)$. We use $\gamma_{\rm i}$ to express the interference power scale, and use $2R_{\rm i}/c$ to represent the interference delay w.r.t. the reference time of a pulse observation interval. Denote the center frequency of the LFM interference that falls in the SAR observation RF band as $f_{\rm i}$, then the observed interference signal can be expressed as
\begin{equation}\label{}
  s_{\rm i}(\eta,\tau)=\gamma_{\rm i}\delta(\eta)h_{\rm i}\left(\tau-\frac{2R_{\rm i}}{c}\right)\exp\left(j2\pi f_{\rm i}\left(\tau-\frac{2R_{\rm i}}{c}\right)\right).
\end{equation}
Here the Dirac function $\delta(\eta)$ is used to indicate that the LFM interference is supported at $\eta=0$. In essence, $s_{\rm i}(\eta,\tau)$ is a one-dimensional signal but embedded in a two-dimensional space. It is worth noting that $\gamma_{\rm i}$ is related to the interference power, transmission loss and the antenna patterns of the two SARs. In the time delay expression $2R_{\rm i}/c$, $R_{\rm i}$ indicates that the interference overlaps with the radar echo of a scatterer located at slant range $R_{\rm i}$.

Suppose that  $s_{\rm i}(\eta,\tau)$ is received by the SAR's receiver without saturation and distortion, the LFM interference in the baseband SAR raw data is the sample of
\begin{equation}\label{}
\begin{split}
 s_{\rm i}(\eta,\tau)=&\gamma_{\rm i}\delta(\eta)h_{\rm i}\left(\tau-\frac{2R_{\rm i}}{c}\right)\exp\left(j2\pi  f_{\rm i,0}\tau\right)\\
 &\cdot\exp\left(-j\frac{4\pi f_{\rm i}R_{\rm i}}{c}\right),
\end{split}
\end{equation}
where $ f_{\rm i,0}=f_{\rm i}-f_0$ is the carrier frequency gap between the passively received interference signal and the actively transmitted radar signal.

To model the image-domain artefact signatures caused by the LFM interference $s_{\rm i}(\eta,\tau)$, in the next subsection we will adopt the aforementioned omega-K algorithm to derive a closed-form expression for the interference response.

\subsection{Analytical Model of Mutual Interference Artefact Based on Omega-K Algorithm}
According to the omega-K algorithm mentioned earlier, the first step for deriving the image-domain response of the interference  is to transform $s_{\rm i}(\eta,\tau)$ into the wavenumber domain. By POSP, the desired two-dimensional signal expression is
\begin{equation}\label{}
  S_{\rm i,1}(f_{\eta},f_{\tau})=\gamma_{\rm i}\operatorname{rect}\left(\frac{f- f_{\rm i,0}}{K_{\rm i}T_{\rm i}}\right)\exp\left(j\theta_1(f_{\eta},f_{\tau})\right),
\end{equation}
where
\begin{equation}\label{}
 \theta_1(f_{\eta},f_{\tau})=-\frac{\pi}{K_{\rm i}}\left(f_{\tau}- f_{\rm i,0}\right)^2-\frac{4\pi R_{\rm i}}{c}(f_{\rm i}+f_{\tau}- f_{\rm i,0}).
\end{equation}
 It can be seen that $S_{\rm i,1}(f_{\eta},f_{\tau})$ is an LFM signal in range frequency and it does not change w.r.t. Doppler frequency.

The second step is to multiply $S_{\rm i,1}(f_{\eta},f_{\tau})$ by $\exp(j\theta_{\rm ref})$. Here the processed Doppler band should be considered, which we denote as ${\rm rect}\big(\frac{f_{\eta}-f_{\eta_{\rm c}}}{B_{\rm p}}\big)$ and $B_{\rm p}$ is the processed pulse repetition interval (PRF). Based on the approximation
\begin{equation}\label{}
\sqrt{(f_0+f_{\tau})^2-\frac{c^2f_{\eta}^2}{4V^2}}
\approx \frac{f_{\tau}}{D(f_{\eta_{\rm c}},V)}+f_0D(f_{\eta},V),
\end{equation}
the resulting signal can be approximated as
\begin{equation}\label{}
\begin{split}
  S_{\rm i,2}(f_{\eta},f_{\tau})= &\gamma_{\rm i}\,{\rm rect}\left(\frac{f_{\eta}-f_{\eta_{\rm c}}}{B_{\rm p}}\right)\operatorname{rect}\left(\frac{f- f_{\rm i,0}}{K_{\rm i}T_{\rm i}}\right)\\
  &\cdot\exp\left(j\theta_2(f_{\eta},f_{\tau})\right),
\end{split}
\end{equation}
where
\begin{equation}\label{}
\begin{split}
 \theta_2(f_{\eta},f_{\tau})=&-\frac{\pi f_{\tau}^2}{K'_{\rm i}}-4\pi\left(\frac{R_{\rm i}}{c}-\frac{R_{\rm ref}}{cD(f_{\eta_{\rm c}},V)}-\frac{f_{\rm i,0}}{K_{\rm i}}\right)f_{\tau}\\
 &+\frac{4\pi f_0R_{\rm ref}}{c}D(f_{\eta},V)-\frac{4\pi R_{\rm i}}{c}(f_{\rm i}- f_{\rm i,0})\\
 &-\frac{\pi f_{\rm i,0}^2}{K_{\rm i}},
\end{split}
\end{equation}
and
\begin{equation}\label{}
  K'_{\rm i}=\frac{K_{\rm i}K_{\rm r}}{K_{\rm r}-K_{\rm i}}.
\end{equation}
Here $K_{\rm r}\neq K_{\rm i}$ is assumed, which is a common case due to different designs between different SAR systems. Under this assumption, $S_{\rm i,2}(f_{\eta},f_{\tau})$ is still an LFM signal in range frequency.

In the third step, we transform the resulting signal into range-Doppler domain. By POSP, the desired signal can be given as
\begin{equation}\label{}
\begin{split}
   S_{\rm i,3}(f_{\eta},{\tau})= & \gamma_{\rm i}\,{\rm rect}\left(\frac{f_{\eta}-f_{\eta_{\rm c}}}{B_{\rm p}}\right){\rm rect}\left(\frac{K'_{\rm i}}{K_{\rm i}T_{\rm i}}\left(\tau-\tau_{\rm i}\right)\right)\\
   &\cdot\exp\left(j\theta_3(f_{\eta},{\tau})\right),
\end{split}
\end{equation}
where
\begin{align}\label{}
  &\tau_{\rm i}=2\left(\frac{R_{\rm i}}{c}-\frac{R_{\rm ref}}{cD(f_{\eta_{\rm c}},V)}-\frac{f_{\rm i,0}}{K_{\rm i}}\right),\\
  &\theta_3(f_{\eta},{\tau})=\pi K'_{\rm i}(\tau-\tau_{\rm i})^2+\frac{4\pi R_{\rm ref}}{c}f_0D(f_{\eta},V)\\\notag
 &\qquad\; -\frac{4\pi R_{\rm i}}{c}(f_{\rm i} - f_{\rm i,0})-\frac{\pi f_{\rm i,0}^2}{K_{\rm i}}.
\end{align}
It is clear that $S_{\rm i,3}(f_{\eta},{\tau})$ is also an LFM signal in this domain. If we consider a particular case of zero Doppler and $f_{\rm i,0}=0$, the LFM interference artefact is centered at $\tau=2(R_{\rm i}-R_{\rm ref})/c$.

As for the case of $K_{\rm r}\neq K_{\rm i}$, $S_{\rm i,3}(f_{\eta},{\tau})$ admits a different form,
\begin{equation}\label{}
\begin{split}
   S_{\rm i,3}(f_{\eta},{\tau})=& \gamma_{\rm i}\left|K_{\rm i}T_{\rm i}\right|{\rm rect}\left(\frac{f_{\eta}-f_{\eta_{\rm c}}}{B_{\rm p}}\right){\rm sinc}\left({K_{\rm i}T_{\rm i}}\left(\tau-\tau_{\rm i}\right)\right)\\
   &\cdot\exp\left(j\theta_3(f_{\eta},{\tau})\right),
\end{split}
\end{equation}
and accordingly, $\theta_3(f_{\eta},{\tau})$ will be
\begin{align}\label{}
  &\theta_3(f_{\eta},{\tau})=\frac{4\pi R_{\rm ref}}{c}f_0D(f_{\eta},V)-\frac{4\pi R_{\rm i}}{c}(f_{\rm i} - f_{\rm i,0})-\frac{\pi f_{\rm i,0}^2}{K_{\rm i}}.
\end{align}
In this case the interference will exist in a form of bright lines in the range-Doppler domain, as we observed during this study. For generality, in the following we will consider the former case, i.e., $K_{\rm r}\neq K_{\rm i}$.

After the fourth step, the phase $\theta_3(f_{\eta},{\tau})$ is changed to
\begin{equation}\label{theta4}
\begin{split}
  \theta_4(f_{\eta},{\tau})=\;&\pi K'_{\rm i}(\tau-\tau_{\rm i})^2+\frac{4\pi R_{\rm 0}}{c}f_0D(f_{\eta},V)\\
  &-\frac{4\pi R_{\rm i}}{c}(f_{\rm i} - f_{\rm i,0})-\frac{\pi f_{\rm i,0}^2}{K_{\rm i}}.
\end{split}
\end{equation}

As the final step, azimuth inverse transform is performed, i.e.,
\begin{equation}\label{H5}
\begin{split}
  s_{\rm i,5}({\eta},{\tau})=\int &\gamma_{\rm i}\,{\rm rect}\left(\frac{f_{\eta}-f_{\eta_{\rm c}}}{B_{\rm p}}\right){\rm rect}\left(\frac{K'_{\rm i}}{K_{\rm i}T_{\rm i}}\left(\tau-\tau_{\rm i}\right)^2\right)\\
  &\cdot\exp\left(j\theta_4(f_{\eta},{\tau})\right)\exp(j2\pi f_{\eta}\eta)df_{\eta}.
\end{split}
\end{equation}
For providing a closed-form solution, we apply POSP once again. As mentioned earlier, the stationary phase points in the integral satisfy the zero-derivative condition, i.e.,
\begin{equation}\label{dtheta}
  \frac{d \theta_4(f_{\eta},{\tau})}{df_{\eta}}+2\pi \eta=0.
\end{equation}
By the definitions of $\theta_4(f_{\eta},{\tau})$ and $D(f_{\eta},V)$ in (\ref{theta4}) and (\ref{dfv}), respectively, (\ref{dtheta}) can be written as
\begin{equation}\label{}
  -\frac{4\pi f_0R_0}{c}\frac{c^2f_{\eta}}{4V^2f^2_0\sqrt{1-\frac{c^2f_{\eta}^2}{4V^2f_0^2}}}+2\pi {\eta}=0.
\end{equation}
From this equation, $f_{\eta}$ is uniquely determined by $\eta$, i.e.,
\begin{equation}\label{}
  f_{\eta}=\frac{2V^2f_0\eta}{c\sqrt{R_0^2+V^2\eta^2}}.
\end{equation}
Inserting this expression into the function in integral (\ref{H5}), we obtain
\begin{equation}\label{h5}
\begin{split}
  s_{\rm i,5}({\eta},{\tau}) = & \,\gamma_{\rm i}\,{\rm rect}\left(\frac{\frac{2V^2f_0\eta}{c\sqrt{R_0^2+V^2\eta^2}}-f_{\eta_{\rm c}}}{B_{\rm p}}\right)\\
  & \cdot {\rm rect}\left(\frac{\tau-\tau_{\rm i}}{\frac{K_{\rm i}}{K'_{\rm i}}T_{\rm i}}\right)\exp(j\theta_5(\eta,\tau)),
\end{split}
\end{equation}
where
\begin{align}\label{}
&\tau_{\rm i}=2\left(\frac{R_{\rm i}}{c}-\frac{R_{\rm ref}}{cD(f_{\eta_{\rm c}},V)}-\frac{f_{\rm i,0}}{K_{\rm i}}\right),\label{tauj}\\
  &\theta_5(\eta,\tau)=\pi K'_{\rm i}\left(\tau-\tau_{\rm i}\right)^2+C\label{theta5}\\
  &\qquad\qquad\;+\frac{4\pi R_0f_0}{c\sqrt{1+\frac{V^2\eta^2}{R_0^2}}}+\frac{4\pi V^2f_0\eta^2}{cR_0\sqrt{1+\frac{V^2\eta^2}{R_0^2}}},\notag\\
  &C=-\frac{4\pi R_{\rm i}}{c}(f_{\rm i} - f_{\rm i,0})-\frac{\pi f_{\rm i,0}^2}{K_{\rm i}}.
\end{align}
Here the phase expression $\theta_5(\eta,\tau)$ actually does not show that $s_{\rm i,5}({\eta},{\tau})$ is LFM in azimuth time. However, it can be well approximated by an LFM signal, as we will show next.

\begin{figure*}[htbp]
  \centering
  \subfigure[$0.6^{\circ}$ squint angle]{\includegraphics[width=5.85cm]{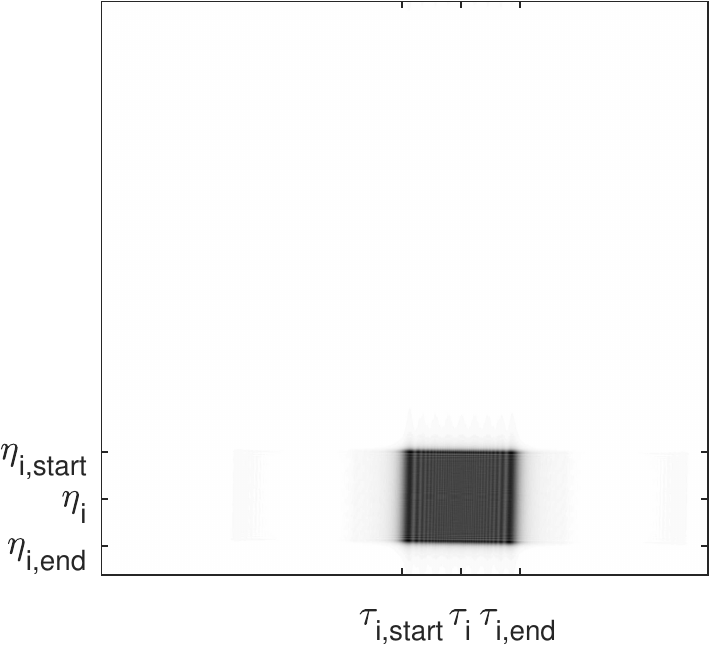}}
    \subfigure[$0.0^{\circ}$ squint angle]{\includegraphics[width=5.85cm]{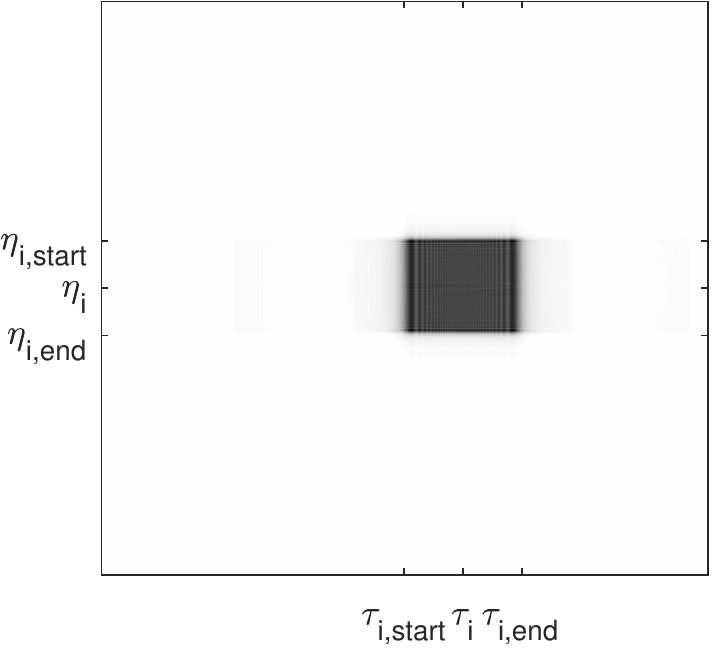}}
     \subfigure[$-0.6^{\circ}$ squint angle]{ \includegraphics[width=5.85cm]{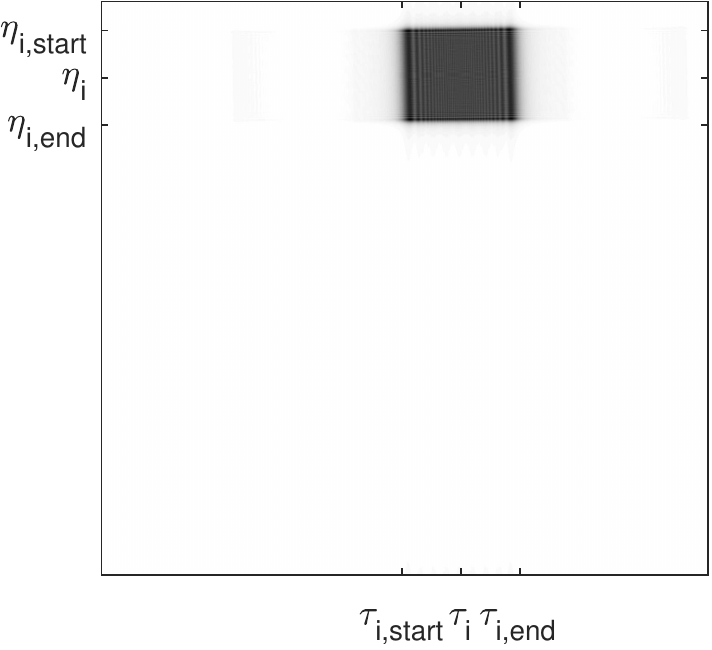}}\\
  \caption{Spatial location, extent and their predictions using (\ref{eta_j})-(\ref{dttau_j}). Predicted azimuth and range locations: $\eta_{\rm i}$ and $\tau_{\rm i}$; predicted azimuth  and range extents: $\Delta\eta_{\rm i}=\eta_{\rm i,start}-\eta_{\rm i,end}$ and $\Delta\tau_{\rm i}=\tau_{\rm i,start}-\tau_{\rm i,end}$. Range is horizontal and azimuth is vertical. Reference point is located at the center of the black box.}\label{sc}
\end{figure*}

In the above expression, $\frac{V^2\eta^2}{R_0^2}$ is much less than $1$. For example, consider a synchronous orbit satellite SAR with effective velocity $7100$ [m/s], scene center slant range $850$ [km] and synthetic aperture time less than $1$ [s], then $\frac{V^2\eta^2}{R_0^2}$  will be less than $1\times 10^{-4}$. Therefore, we can apply the following approximation,
\begin{equation}\label{}
\frac{1}{\sqrt{1+\frac{V^2\eta^2}{R_0^2}}}\approx 1-\frac{V^2\eta^2}{2R_0^2}.
\end{equation}
Then the last two terms in (\ref{theta5}) can be approximated as
\begin{equation}\label{}
\begin{split}
&\frac{4\pi R_0f_0}{c\sqrt{1+\frac{V^2\eta^2}{R_0^2}}}+\frac{4\pi V^2f_0\eta^2}{cR_0\sqrt{1+\frac{V^2\eta^2}{R_0^2}}}\\
 &\approx \pi K_{\rm a}\eta^2\left(1-\frac{V^2\eta^2}{2R_0^2}\right)+\frac{4\pi R_0f_0}{c}\\
 &\approx \pi K_{\rm a}\eta^2+\frac{4\pi R_0f_0}{c},
\end{split}
\end{equation}
where $K_{\rm a}=\frac{2V^2 f_0}{cR_0}$ is the Doppler FM rate.
Then (\ref{theta5}) can be approximated as
\begin{equation}\label{}
\begin{split}
  &\theta_5(\eta,\tau)=\pi K'_{\rm i}\left(\tau-\tau_{\rm i}\right)^2+\pi K_{\rm a}\eta^2+\frac{4\pi R_0f_0}{c}+C.
\end{split}
\end{equation}
Now it is clear that the phase of the interference response is quadratic in both range and azimuth times, so we can view $s_{\rm i,5}({\eta},{\tau})$ as a two-dimensional LFM signal under mild approximations.

\section{Interpretations of Mutual Interference Artefact Characteristics}

In this section we are going to analyse the spatial and spectral characteristics of the LFM interference artefact, based on the above theoretical model as well as validations using numerical simulations.

\subsection{Spatial Characteristics}

1) \emph{Azimuth Location}. According to (\ref{h5}), the azimuth location, which we denote as $\eta_{\rm i}$, satisfies
\begin{equation}\label{}
  \frac{2V^2f_0\eta_{\rm i}}{c\sqrt{R_0^2+V^2\eta_{\rm i}^2}}=f_{\eta_{\rm c}}.
\end{equation}
A simple algebra operation yields
\begin{equation}\label{eta_j}
  \eta_{\rm i}=\frac{f_{\eta_{\rm c}}}{K_{\rm a}\sqrt{1-\frac{c^2f_{\eta_{\rm c}}^2}{4V^2f_0^2}}}.
\end{equation}
This relation indicates that the interference artefact position is approximately linear in $f_{\eta_{\rm c}}$ for small squint angle cases.
  Since $K_{\rm a}$ decreases with $R_0$, we can also find that the absolute values of interference position $|\eta_{\rm i}|$ increases with $R_0$ (for a single artefact) if $f_{\eta_{\rm c}}\neq 0$. For a large Doppler centroid or squint angle, this behavior is more significant. This can be seen from the first S-1 quick-look image in Fig. \ref{ex}, where the burst edges have larger squint angles. In our simulation that will be presented later, one can zoom in the simulated images to find this behavior.

2) \emph{Range Location}. From  (\ref{h5}), the range location is $\tau_{\rm i}$ as defined in (\ref{tauj}). Suppose that $f_{\rm i,0}=0$ or it is ignorable and $f_{\eta_{\rm c}}=0$, then $\tau_{\rm i}$ is located at $2(R_{\rm i}-R_{\rm ref})/c$, or equivalently, the interference artefact has a distance $R_{\rm i}-R_{\rm ref}$ w.r.t. the reference slant range. In general cases, the distance increases with $|f_{\eta_{\rm c}}|$.

 3) \emph{Azimuth Extent}. This spatial characteristic is determined by the first rectangular function in (\ref{h5}), i.e., the azimuth extent is
 \begin{equation}\label{}
 \begin{split}
   \Delta\eta_{\rm i} & =\eta_{\rm i,end}-\eta_{\rm i,start}\\
   & = \frac{f_{\eta_{\rm c}}+\frac{B_{\rm p}}{2}}{K_{\rm a}\sqrt{1-\frac{c^2(f_{\eta_{\rm c}}+\frac{B_{\rm p}}{2})^2}{4V^2f_0^2}}}- \frac{f_{\eta_{\rm c}}-\frac{B_{\rm p}}{2}}{K_{\rm a}\sqrt{1-\frac{c^2(f_{\eta_{\rm c}}-\frac{B_{\rm p}}{2})^2}{4V^2f_0^2}}}.
 \end{split}
 \end{equation}
 Using the approximation
 \begin{equation}\label{}
   \sqrt{1-\frac{c^2(f_{\eta_{\rm c}}\pm \frac{B_{\rm p}}{2})^2}{4V^2f_0^2}}\approx 1+\frac{c^2f^2_{\eta_c}}{8V^2f_0^2},
 \end{equation}
 the azimuth extent can be expressed as
 \begin{equation}\label{dteta_j}
      \Delta\eta_{\rm i} \approx \frac{B_{\rm p}}{K_{\rm a}}+\frac{c^2f^2_{\eta_c}B_{\rm p}}{8K_{\rm a}V^2f_0^2}\approx \frac{B_{\rm p}}{K_{\rm a}}.
 \end{equation}
 If we ignore the second term (for the case of small Doppler centroid or low squint angle), we will simply have $\Delta\eta_{\rm i} \approx \frac{B_{\rm p}}{K_{\rm a}}$. If we further consider, in particular, a stripmap case where $B_{\rm p}$, for example, is $1.2$ times of the azimuth bandwidth, then the interference artefact's azimuth extent is about $1.2$ times of the synthetic aperture time.

 4) \emph{Range Extent}. By the second rectangular function in (\ref{h5}), the range extent of the interference artefact is
 \begin{equation}\label{dttau_j}
   \Delta\tau_{\rm i}=\left|\frac{K_{\rm i}}{K_{\rm i}'}\right|T_{\rm i}=\left|\frac{K_{\rm r}-K_{\rm i}}{K_{\rm r}}\right|T_{\rm i}.
 \end{equation}
 According to this relation, the range extent increases as $K_{\rm i}$ deviates from $K_r$. It implies that if two spaceborne SARs adopt similar range-time FM rates, then the impact of their single interference artefact is small in terms of range extent. On the other hand,  if the range-time FM rates are much different, the interference influence would be more severe (if the LFM interference happens).

\begin{figure}
  \centering
  \includegraphics[width=6.5cm]{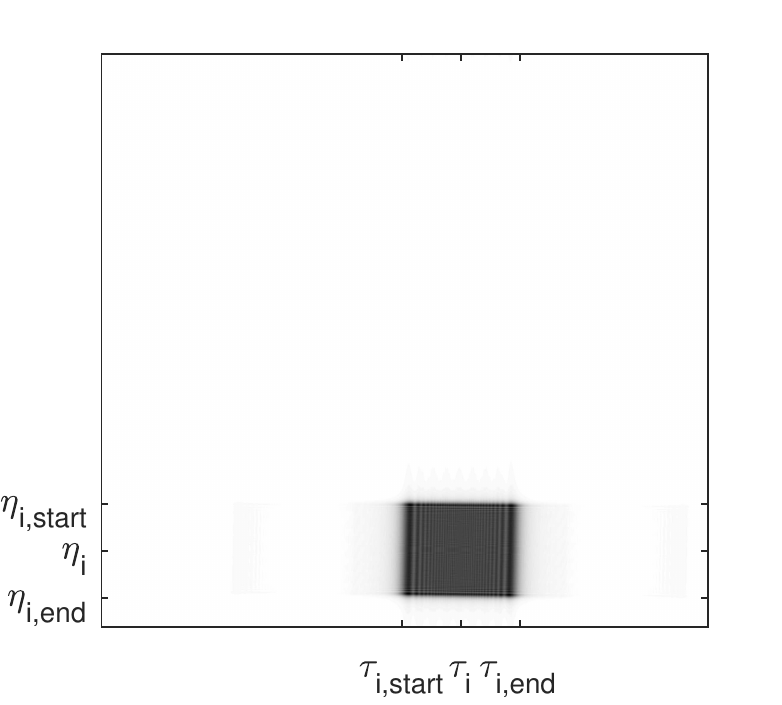}
  \caption{Prediction of spatial extent and location of an LFM interference artefact that is generated via the chirp scaling algorithm. Squint angle is $0.6^{\circ}$. This result coincides with the first image in Fig. \ref{sc}.}\label{csa}
\end{figure}

To verify the above theoretical characterization of interference spatial extent and location, we perform a simple simulation. The involved simulation parameters are as follows: $f_0=5.4 \times10^9$ [Hz], $K_{\rm r}=5\times10^{11}$ [Hz/s], $K_{\rm i}=-2.5\times10^{11}$ [Hz/s],  $V=7100$ [m/s], $T_{\rm i}=1.65\times10^{-5}$ [s], $T=4\times10^{-5}$ [s], $B_{\rm p}=1200$ [Hz], $R_{\rm ref}=850$ [km], and the squint angle is considered to be $0.6^{\circ}$, $0.0^{\circ}$ and $-0.6^{\circ}$, respectively. The interference artefacts calculated in the three squint angle settings are shown in Fig. \ref{sc}. The simulated azimuth and range sampling points is $4096\times 2048$. The predicted spatial extents and locations of the interference artefacts via  (\ref{eta_j})-(\ref{dttau_j}) are marked using
\begin{equation}\label{}
  \eta_{\rm i,start},\;\eta_{\rm i,end},\;\eta_{\rm i},\;\;\tau_{\rm i,start},\;\tau_{\rm i,end},\;\tau_{\rm i}.
\end{equation}
The results show that the predictions well match the simulated LFM interference artefacts. Furthermore, it is notable that using a different focusing algorithm still leads to very similar characteristics, as the chirp scaling algorithm-based simulation in Fig. \ref{csa} shows.

\subsection{Spectral Characteristics}

\begin{figure}
  \centering
  \includegraphics[width=8.8cm]{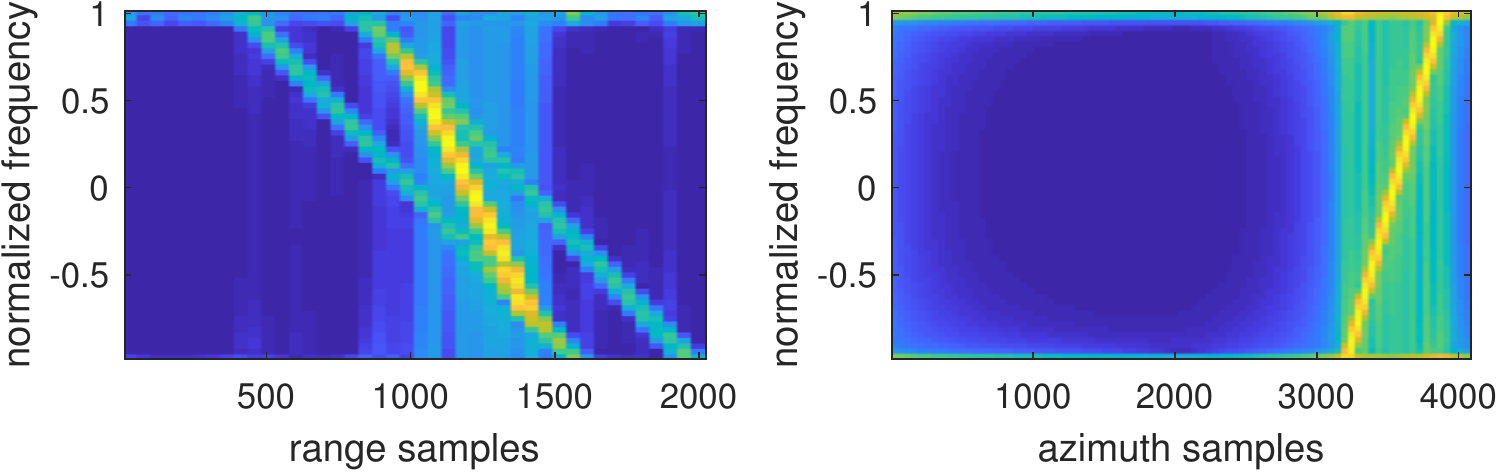}\\\vspace{0.15cm}
    \includegraphics[width=8.8cm]{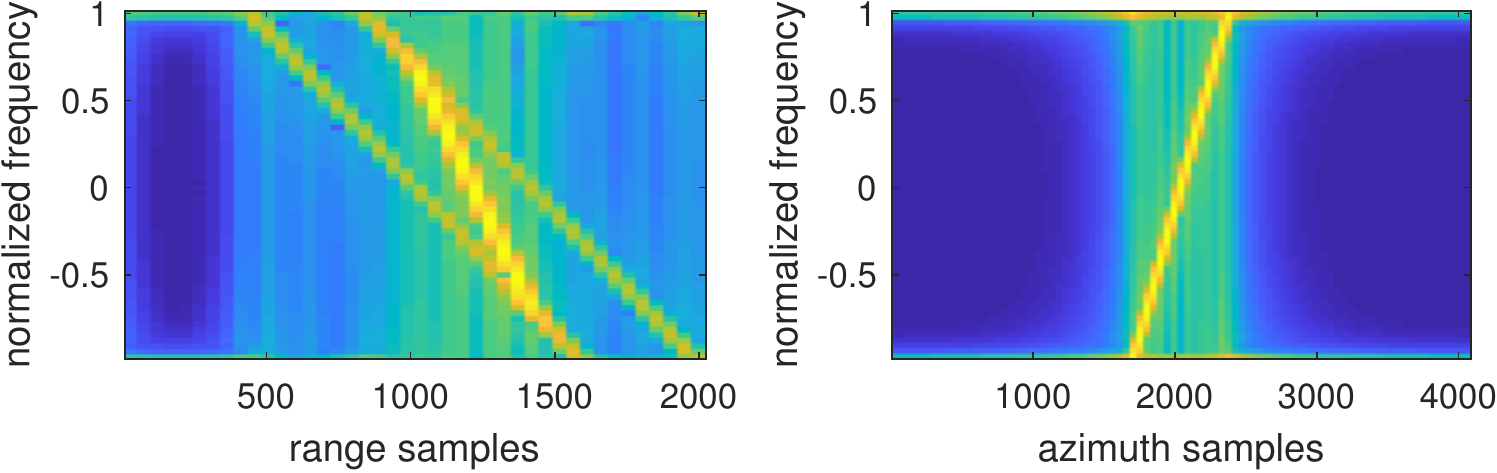}\\\vspace{0.15cm}
      \includegraphics[width=8.8cm]{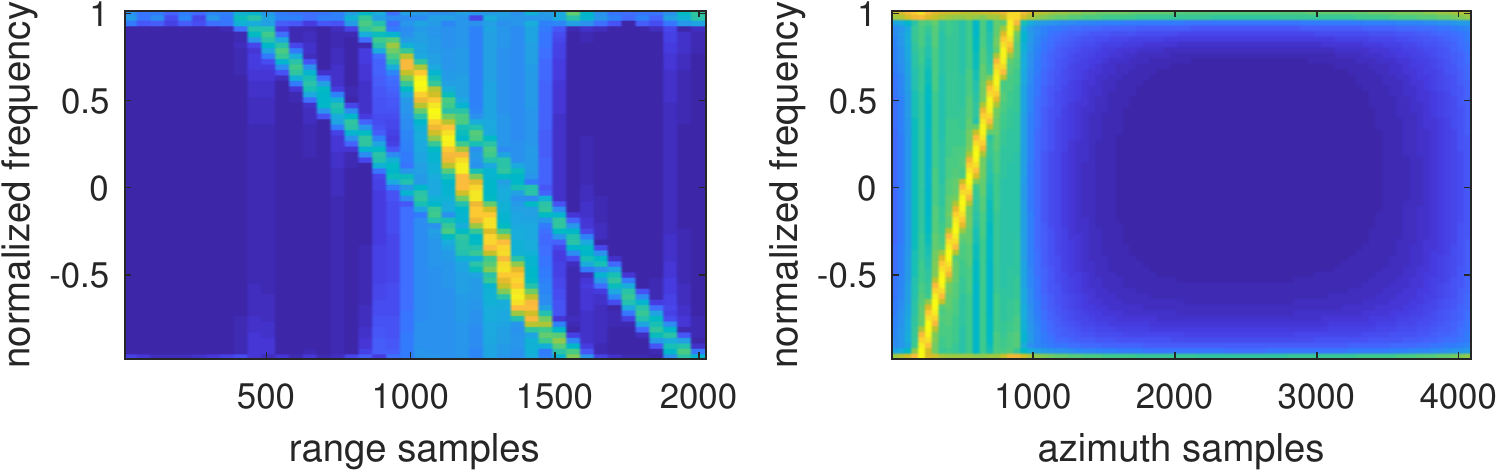}
  \caption{Time-frequency diagrams in range and azimuth (obtained via short-time Fourier transform). }\label{tfd}
\end{figure}

For spectral characteristics of the interference artefact, the notable point is the two-dimensional LFM property mentioned earlier. In the azimuth direction, the FM rate is uniquely determined by the SAR imaging geometry, and it does not change with the interference signal. In the range direction, the interference artefact's FM rate depends on the range-time FM rates of the both SARs. We plot the short-time Fourier transform (STFT) in Fig. \ref{tfd} for visualizing the spectral characteristics of the interference artefact.

Since in both directions the interference artefact is LFM signal, a feasible interference mitigation technique is to first apply STFT and then use notch filtering to cut off the interference artefact in time-frequency domain, which is similar to raw data-domain time-varying filtering \cite{zhang2011,Tao2016,8282759}. However, this solution has two drawbacks. First, STFT has coarse frequency resolutions and thus wide side lobes, which makes the efficient removal of the interference artefact a tricky problem. Second, if the FM rates are not known a priori, they need to be estimated from the image data. If the FM rate estimations are not accurate, then the interference mitigation performance will unavoidably decrease. To better remove the interference artefact, in the following we will propose to use two simple yet efficient image-domain interference mitigation techniques, which starts with the low rank modeling for the interference artefact.
\section{Low Rank Property of Mutual Interference Artefact}

The rank of a matrix is the dimension of its column or row subspace. A low-rank matrix lies on a low-dimensional subspace, and thus has structural information in some sense. This section will present a foundation for modeling the low rank property of the LFM mutual interference artefact, to provide a path to efficient image-domain interference mitigation techniques that will be introduced in the next section.

\subsection{Rank-$1$ Approximation}

\begin{figure*}
  \centering
  \includegraphics[width=16.5cm]{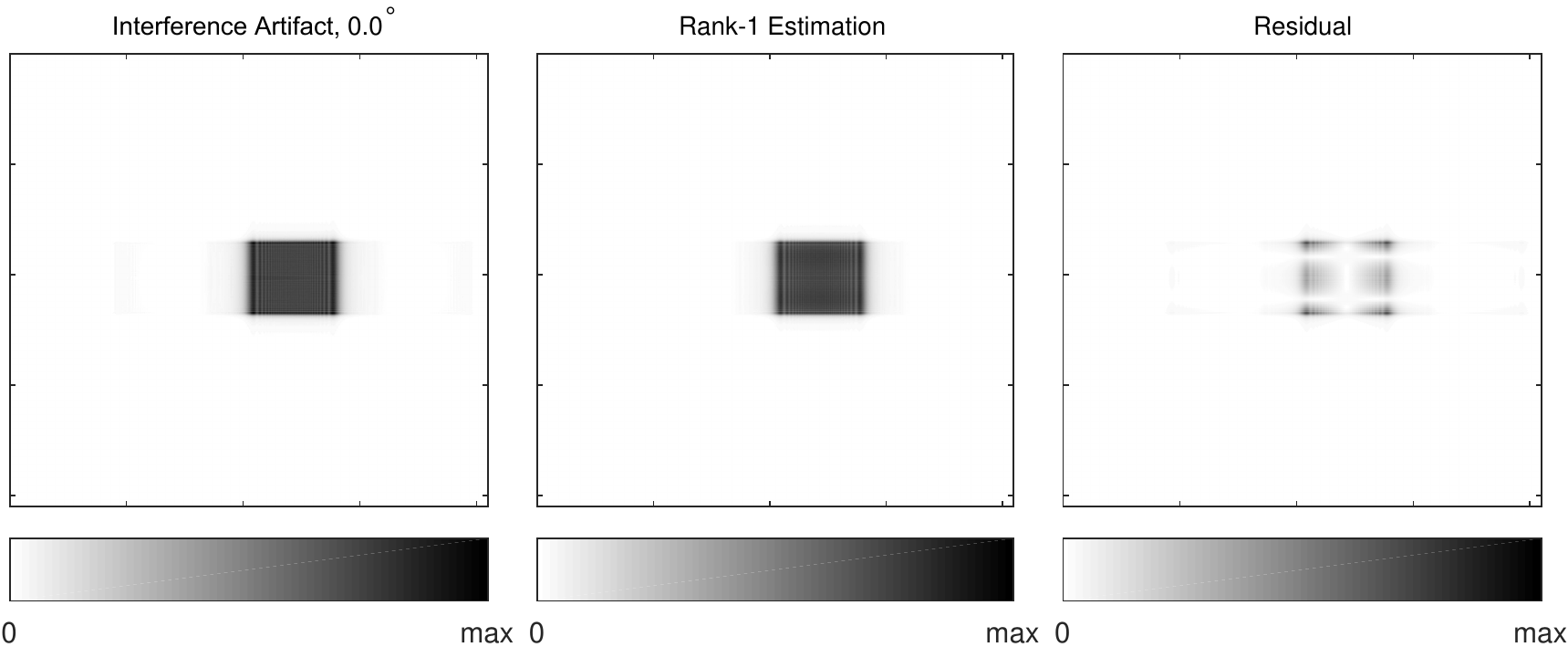}
  \caption{Rank-1 estimation of the simulated LFM interference artefact with zero squint angle. The rank-$1$ approximation error is $0.069$. `max' denotes the maximum intensity over all pixels.}\label{r1}
\end{figure*}
\begin{figure}[t]
  \centering
    \quad\includegraphics[width=5.8cm]{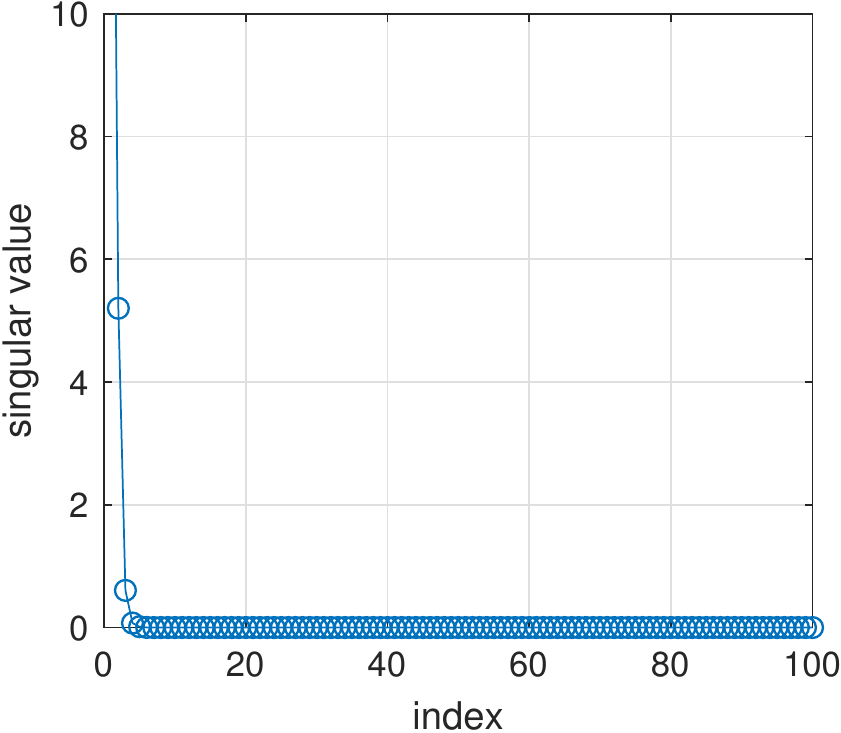}
      \includegraphics[width=6.2cm]{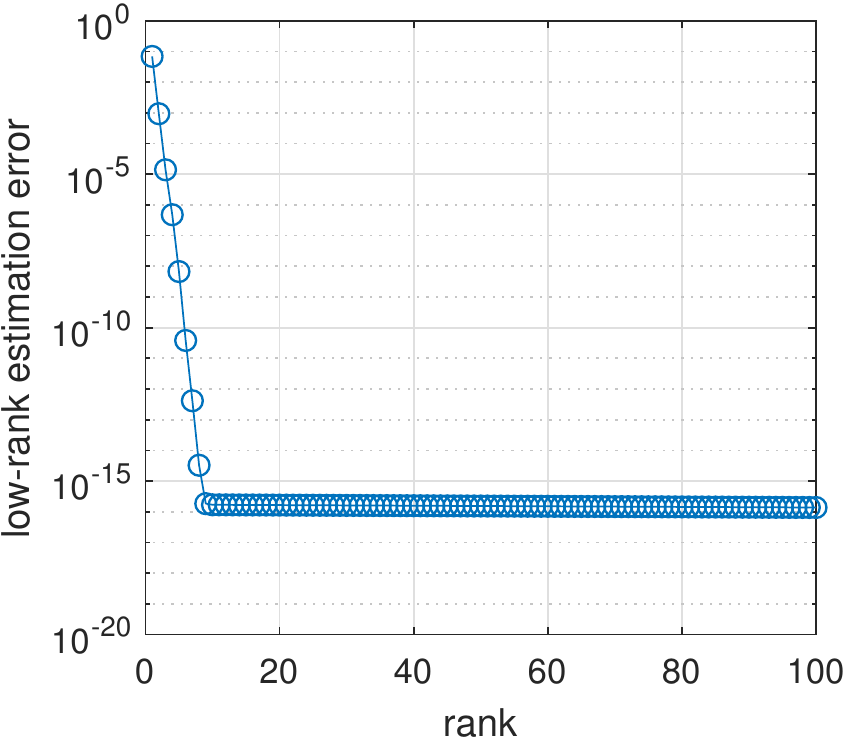}\quad
  \caption{Top $100$ in $2048$ singular values and low rank estimation errors in the case of zero squint angle.}\label{rerr0}
\end{figure}

For a two-dimensional signal with decoupled variables, e.g., $f(x_1,x_2)=\alpha u(x_1)v(x_2)$ where $\alpha$ is a scalar, and $u(\cdot)$, $v(\cdot)$ are one-dimensional signals, one can easily observe that the sample matrix of $f(x_1,x_2)$ is rank-one. In our topic, the interference artefact, $s_{\rm i,5}(\eta,\tau)$, is also rank-$1$ if we adopt appropriate approximations. Specifically, we adopt the following approximation,
\begin{align}\label{}
&K_{\rm a}=\frac{2V^2f_0}{cR_0}\approx\frac{2V^2f_0}{cR_{\rm ref}}:={K_{\rm a,ref}},\\
&\frac{2V^2f_0\eta}{c\sqrt{R_0^2+V^2\eta^2}}\approx \frac{\eta}{K_{\rm a}}\approx \frac{\eta}{K_{\rm a,ref}}.
\end{align}
Then accordingly, $s_{\rm i,5}(\eta,\tau)$ can be approximated as a rank-$1$ signal,
\begin{equation}\label{}
  s_{\rm i,5}(\eta,\tau)\approx \gamma_{\rm i} u_{\rm i}(\eta)v_{\rm i}(\tau),
\end{equation}
where
\begin{align}\label{}
  &u_{\rm i}(\eta)={\rm rect}\left(\frac{K_{\rm a,ref}\eta-f_{\eta}}{B_{\rm p}}\right)\exp\left(j\pi K_{\rm a,ref}\eta^2+jC\right)\\
  &v_{\rm i}(\tau)={\rm rect}\left(\frac{\tau-\tau_{\rm i}}{\frac{K_{\rm i}}{K'_{\rm i}}T_{\rm i}}\right)\exp\left(j\pi K'_{\rm i}\left(\tau-\tau_{\rm i}\right)^2+j\frac{4\pi R_0f_0}{c}\right)
\end{align}

The above rank-$1$ approximation is accurate when the synthetic aperture time and  Doppler centroid are small. For example, Figs. \ref{r1} and \ref{rerr0} show a case of zero squint angle, where rank-$1$ estimation is numerically investigated. In the simulated case we consider an image support of $4096\times 2048$ (azimuth $\times$ range) pixels, and the singular values of the  artefact in the image is shown in Fig. \ref{rerr0}. The rank-$1$ estimation in this simulated case has merely $0.069$ relative approximation error measured in squared Frobenius norm. Specifically, the relative approximation error we adopted is defined as
\begin{equation}\label{}
  \frac{\|{\bf J-J}_{\rm est}\|_{\rm F}^2}{\|{\bf J}\|_{\rm F}^2},
\end{equation}
where $\bf J$ is the interference artefact matrix and ${\bf J}_{\rm est}$ is its estimation.

However, in some observation modes, for example, TOPS, the Doppler centroid can be significantly large due to antenna rotation. Therefore, rank-$1$ approximation may not be a good choice.  Next we will show that the rank-$1$ approximation can be efficiently relaxed via low rank modeling.

\subsection{Low Rank Relaxation}

\begin{figure*}
  \centering
{\includegraphics[width=17cm]{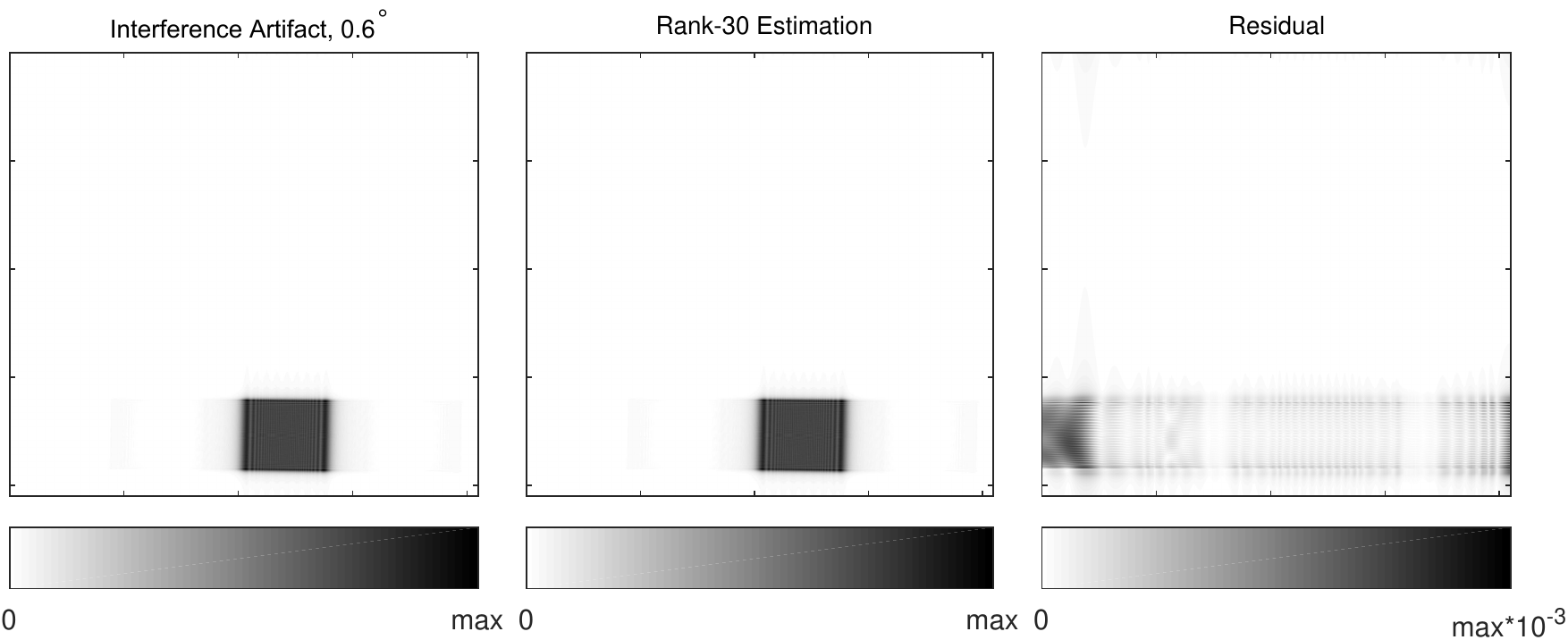}}
  \caption{Rank-30 estimation of the simulated LFM interference artefact with  $0.6^{\circ}$ squint angle. The estimation error is $3.86*10^{-7}$. `max' denotes the maximum intensity over all pixels.}\label{r06}
\end{figure*}

The rank-$1$ approximation error, as one can find, is due to ignoring high-order terms of the two-dimensional artefact. In the aforementioned rank-$1$ approximation, only quadratic phase terms are considered, and thus the approximation accuracy is limited. If we keep more high-order terms, the artefact can be better approximated. To show this point, let us denote the true artefact (i.e., without approximation) as $\widetilde{s}_{\rm i}(\eta,\tau)$. Although we do not know its closed-form expression, we can still decompose it via two-dimensional order-$n$ Taylor expansion,
\begin{equation}\label{}
\widetilde{s}_{\rm i}(\eta,\tau) \approx  \widetilde{s}_{\rm i}(\eta_{\rm i},\tau_{\rm i})+\sum_{p=1}^{n}\left(\eta\frac{\partial }{\eta}+\tau\frac{\partial}{\tau}\right)^p \widetilde{s}_{\rm i}(\eta_{\rm i},\tau_{\rm i}),
\end{equation}
where the second term can be further written as
\begin{equation}\label{}
  \sum_{p=1}^{n}\sum_{q=0}^{p}\binom{p}{q}\left(\eta-\eta_{\rm i}\right)^q \left(\tau-\tau_{\rm i}\right)^{p-q}\frac{\partial^p \widetilde{s}_{\rm i}(\eta,\tau)}{\partial\tau^q\partial \tau^{p-q}}\bigg |_{(\eta_{\rm i},\tau_{\rm i})}.
\end{equation}
We can find that the above equation is the summation of a sequence of range-azimuth decoupled functions that admits the following forms,
  \begin{equation}\label{}
    C_k u_k(\eta) v_k(\tau), \quad k=1,\cdots,K,
  \end{equation}
where $K=\sum_{p=1}^{n}\sum_{q=0}^{p}\binom{p}{q}$.
 Since $C_k u_k(\eta) v_k(\tau)$ is rank-$1$, as it is range-azimuth decoupled, the order-$n$ Taylor expansion has a rank not larger than $K$. For example, for order-$3$ and -$4$ Taylor expansions, the rank are not larger than $14$ and $30$, respectively. The above order-$n$ Taylor expansion can also be performed on the phase of $\widetilde{s}_{\rm i}(\eta,\tau)$ only, to use a larger $n$ for keeping high-order phase terms. In a nutshell, we can relax the rank-$1$ approximation via low rank modeling to improve approximation accuracy.

 \begin{figure}
  \centering
    \quad\includegraphics[width=5.8cm]{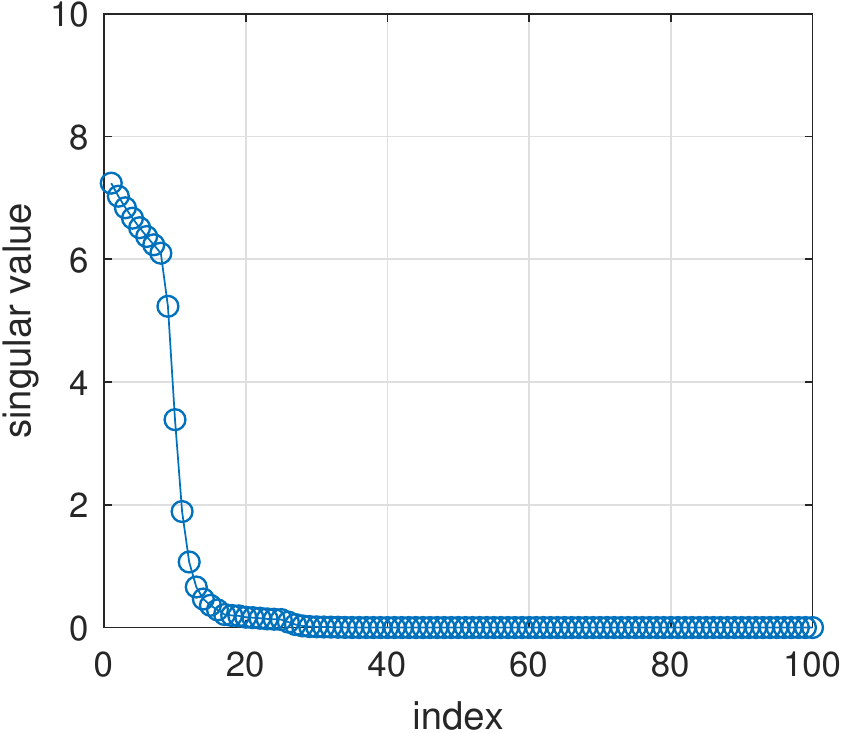}
      \includegraphics[width=6.2cm]{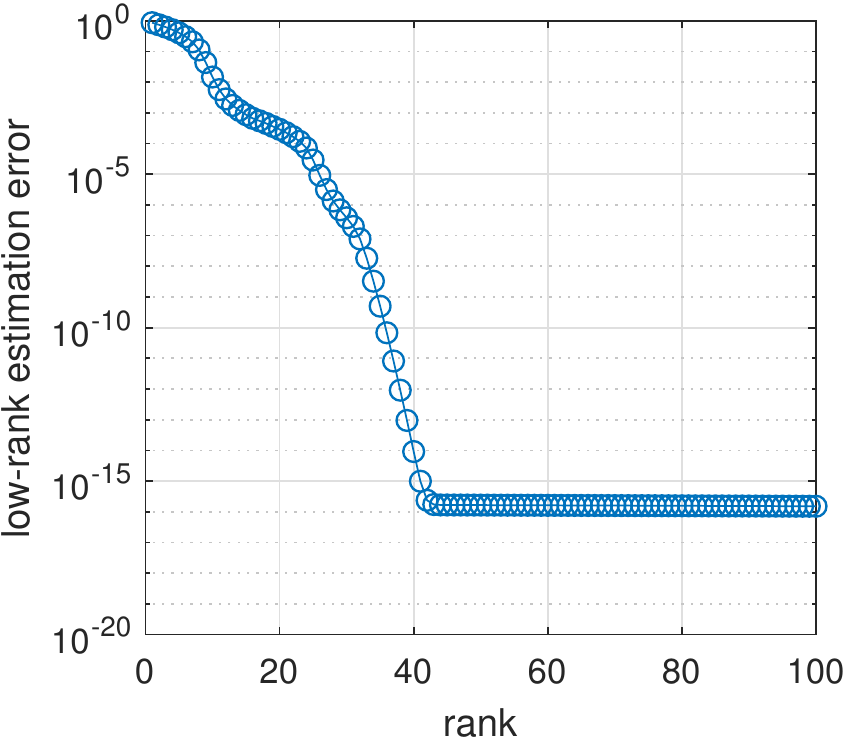}\quad
  \caption{Top $100$ in $2048$ singular values and low rank estimation errors in the case of $0.6^{\circ}$ squint angle. }\label{rerr06}
\end{figure}

To study the accuracy, we simulated a case of $0.6^{\circ}$ squint angle, and the low rank estimation accuracy is evaluated in Figs.  \ref{rerr06} and \ref{r06}. In the simulation the rank-$30$ estimation achieves an approximation error of $3.86\times10^{-7}$, which shows a high accuracy of low rank approximation.

The low rank model acts as an agent for prior information of the interference artefact, and thus can provide a powerful tool towards efficient interference mitigation techniques, as will be presented and validated in the next two sections.

\section{Low Rank-Based Mutual Interference Artefact Mitigation}

Based on the previous low rank modeling, this section briefly introduces two techniques for the mitigation of LFM interference artefact in image domain, i.e., PCA and RPCA.

\subsection{PCA}

PCA tries to find a low-dimension component that best fits a matrix in Frobenius norm. It is widely used in statistics, signal processing, and machine learning. PCA is originally a statistical tool for performing dimensionality reduction. In this paper, however, we will use it as a denoising approach, where interference artefacts are taken as principal components whereas the underlying SAR image is treated as noise. We use PCA to estimate the principal components, and then subtract them to mitigate interference artefacts.

Denote the image patch that contains interference artefacts as
\begin{equation}\label{}
  \bf Y=J+I,
\end{equation}
where $\bf J$ is the artefacts component and $\bf I$ is the underlying SAR image.
The associated PCA optimization problem is
\begin{equation}\label{pca}
  \begin{array}{rl}
    {\bf J}^{\sharp}=\mathop{\rm minimize}\limits_{\bf J} \;\;& \|{\bf J-Y}\|_{\rm F}^2 \\
    {\rm subject\; to} & {\rm rank}({\bf J})\le K.
  \end{array}
\end{equation}
This problem can be solved via $K$ singular value decomposition ($K$-SVD):
\begin{align}\label{}
  &[{\bf U},{\bm\Sigma},{\bf V}]={\rm SVD}({\bf Y})\\
  &{\bf J}^{\sharp}={\bf U}{\bm\Sigma}_K{\bf V}^{\rm H},
\end{align}
where ${\bm\Sigma}_K$ keeps $K$ largest values on its primary diagonal with other elements being zeroed, and $(\cdot)^{\rm H}$ denotes the Hermitian conjugate operation.
Next, the underlying image can be estimated by subtracting the estimated interference component from the original corrupted image data, i.e.,
\begin{equation}\label{}
  {\bf I}^{\sharp}={\bf Y}-{\bf J}^{\sharp}.
\end{equation}

It is important to note that although the classical PCA is effective against the presence of small Gaussian noise, it is highly sensitive to sparse outliers of high magnitude, even they occupy only a small fraction of the matrix. The reason is that a sparse matrix is also low-rank, which may be mistakenly identified by PCA as principal components if its entries have strong magnitudes. To overcome this drawback, we next introduce RPCA to achieve improved interference mitigation performance.

\subsection{RPCA}

Robust PCA tries to decompose a matrix into a low-rank matrix and a sparse one using optimization methods. It has some applications in SAR in terms of RFI suppression\cite{Nguyen2012Robust,SuNarrow2017,HuangNarr2018}, interferometric outlier removal \cite{Kang2018}, moving target indication\cite{7254118,7994703} and so on. RPCA adopts an optimization method called principal component pursuit, i.e.,
\begin{equation}\label{rpca}
  \begin{array}{ll}
    \mathop{\rm{minimize}}\limits_{\bf J,I} & \|{\bf J}\|_*+\mu \|{\bf I}\|_1\\
    \rm{ subject\; to} & \bf J+I=Y,
  \end{array}
\end{equation}
where $\|\cdot\|_*$ is the matrix nuclear norm (i.e., sum of singular values) to introduce low rank, $\|\cdot\|_1$ is the matrix $\ell_1$ norm to introduce sparsity, and $\mu$ is a regularization parameter to balance the low rank and sparsity. This parameter is commonly chosen as ${1}/{\sqrt{\max(N_a,N_r)}}$, where $N_a$ and $N_r$ are the sizes of $\bf Y$ in azimuth and range, respectively. The above problem can be iteratively solved by alternating optimization incorporated with augmented Lagrange method (ALM), i.e.,
\begin{align}\label{}
  & {\bf J}^{k+1}=\mathop{\rm argmin}\limits_{\bf J} \;\|{\bf J}\|_*+\xi_k({\bf Y-J}-{\bf I}^k)\\
  &\hspace{2.4cm}+\frac{\rho}{2}\|{\bf Y-J}-{\bf I}^k\|_{\rm F}^2,\notag\\
  & {\bf I}^{k+1}=\mathop{\rm argmin}_{\bf I} \;\mu\|{\bf I}\|_1+\xi_k({\bf Y-J}^{k+1}-{\bf I})\\
  &\hspace{2.4cm}+\frac{\rho}{2}\|{\bf Y-J}^{k+1}-{\bf I}\|_{\rm F}^2,\notag\\
  &\xi_{k+1}=\xi_k+\rho ({\bf Y-J}^{k+1}-{\bf I}^{k+1}).
\end{align}
The first subproblem can be solved via singular value soft-thresholding and the second one can be solved by merely element-wise soft-thresholding. Since ALM converges to the optimal solution of a convex problem as $\rho \rightarrow +\infty$, the above algorithm does not require complicated parameter tuning.

Next, let us provide some processing examples with S-1 data.

\section{Examples with S-1 Data}

The S-1 MPC has reported a number of  S-1 IW  SLC images corrupted by spaceborne SAR mutual interference from known and unknown satellites \cite{s1mpc,s1mpc2}.
In this section, we will use the aforementioned techniques to perform experiments on interference-corrupted S-1 IW images, to show that the radiometric artefacts caused by LFM interference can be efficiently removed in image domain based on the proposed low rank model.

\subsection{TFD of LFM Mutual Interference Artefact  in S-1 TOPS Mode}
\begin{figure}[t]
  \centering
  \includegraphics[width=8.76cm]{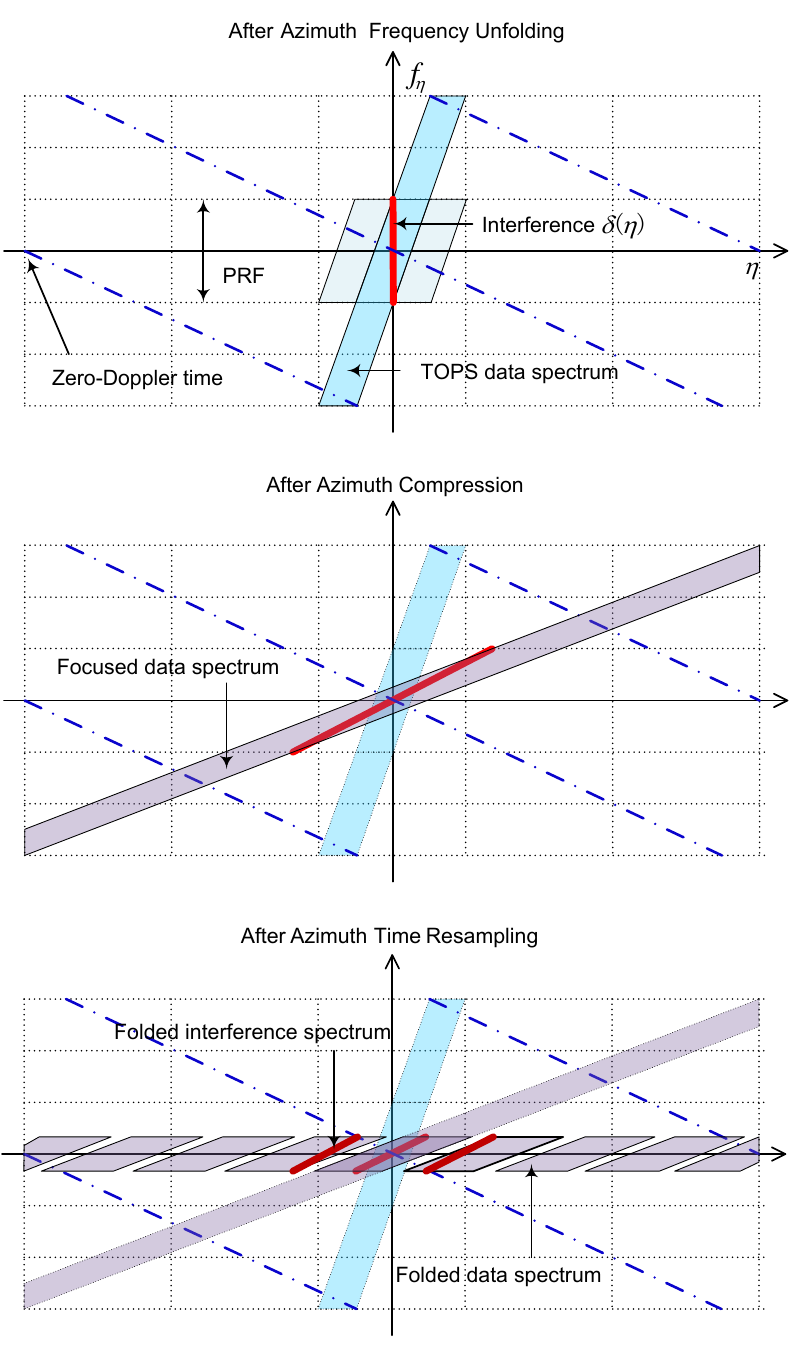}
  \caption{Azimuth TFD illustrations for LFM interference in TOPS mode. Some intermediate steps like azimuth time unfolding and resampling are not shown in the above TFDs for simplicity. }\label{tops}
\end{figure}

\begin{figure}[htbp]
  \centering
 \subfigure{\includegraphics[width=8.75cm]{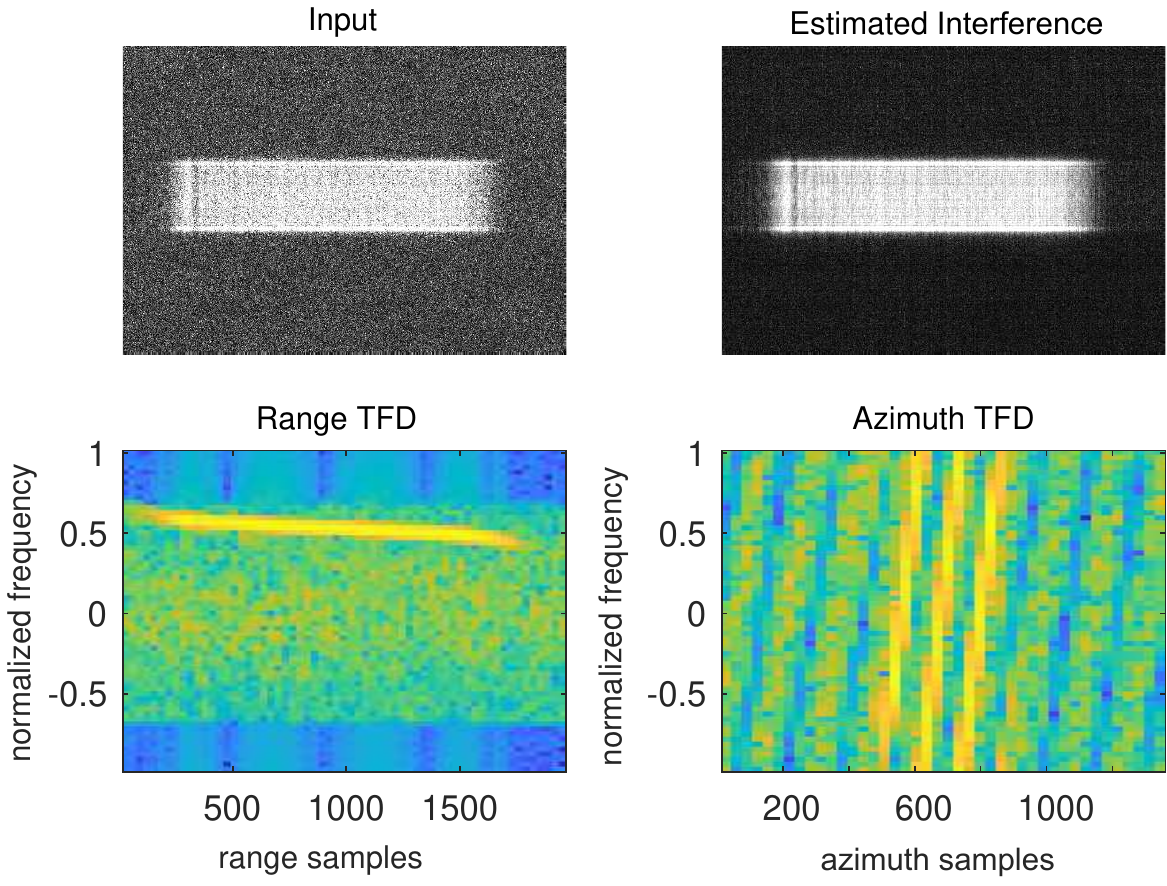}}
  \caption{An example of single LFM interference artefact and its spectra, observed by S-1.  The estimated interference artefact is obtained via PCA with $K=40$.}\label{ice_sub1}
\end{figure}

\begin{figure}[htbp]
  \centering
    \subfigure{\includegraphics[width=8.75cm]{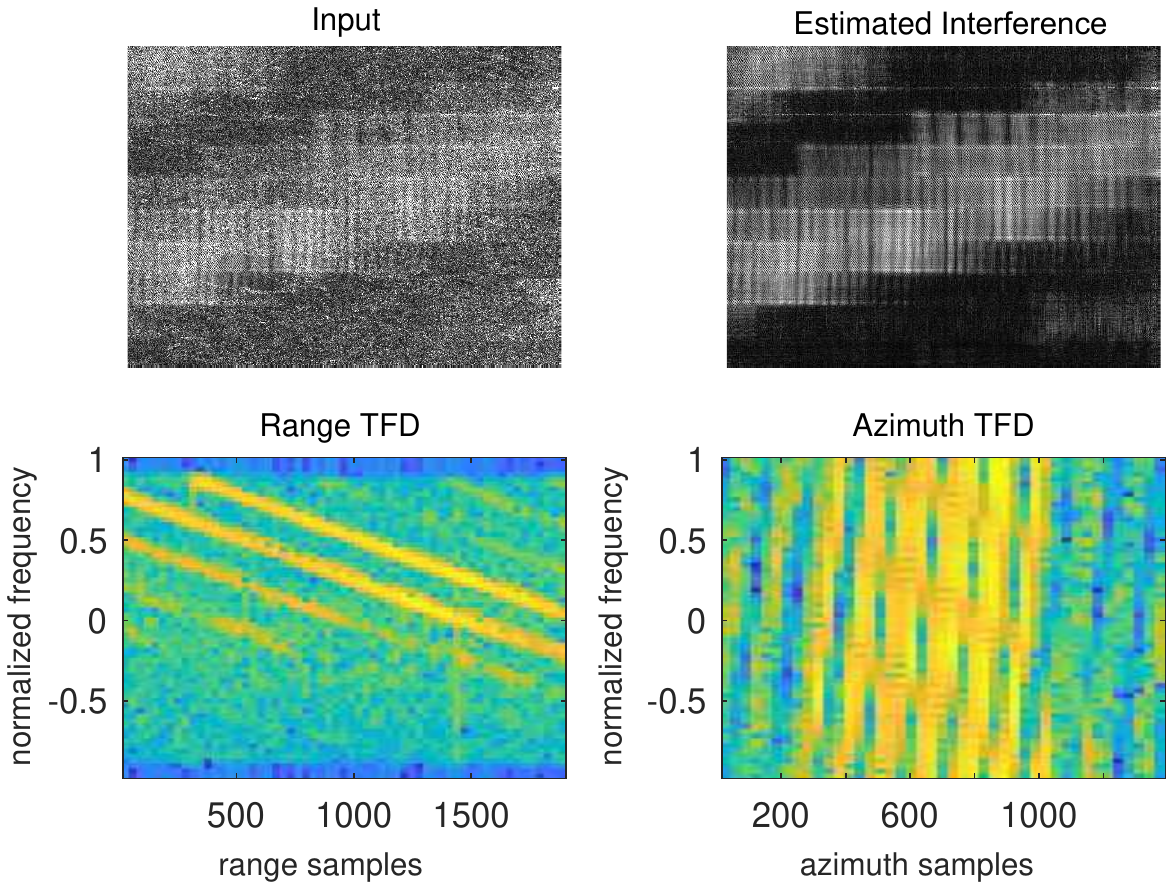}}
  \caption{An example of multiple LFM interference artefacts and their spectrogram, observed by S-1.  The estimated interference artefacts are obtained via PCA with $K=40$.}\label{ice_sub2}
\end{figure}

\begin{figure}[htbp]
  \centering
  \includegraphics[width=8.6cm]{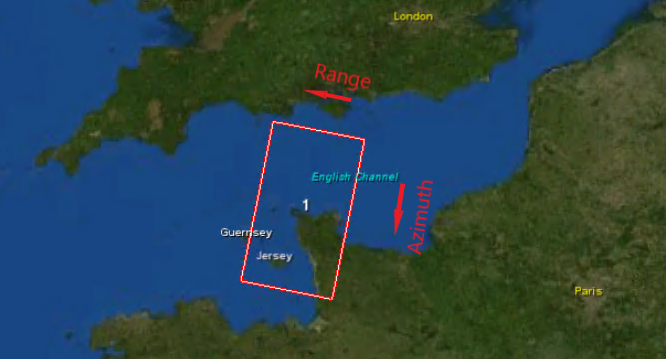}
  \caption{The footprint of the S-1 IW1 image.}\label{s1_france}
\end{figure}

The S-1 IW works in TOPS mode \cite{s1a}, of which the image focusing process is different form the standard stripmap processing. The main difference is in the azimuth direction, where TOPS focusing adopts additional steps in azimuth to deal with time- and frequency-domain folding\cite{s1a,1677745}. This is slightly different from the previous simulated stripmap examples. To further explain the folded interference artefact's azimuth spectrum, we provide three time-frequency diagrams (TFD) in TOPS processing, as shown in Fig. \ref{tops}.

The first TFD in Fig. \ref{tops} shows the azimuth spectrum after azimuth frequency unfolding and resampling. The red vertical line indicates the interference spectrum, which occupies the full PRF since it is a Dirac function in azimuth time. The second TFD shows the spectrum after azimuth focusing, where the spectrum of each target is processed to be vertical and moved to its zero-Doppler time. Accordingly, the interference spectrum now has a slope in the time-frequency domain, which is linear frequency modulated. In the third TFD, the focused data is down sampled with an appropriate rate to obtain the final azimuth pixel spacing, which is slightly higher than the instantaneous bandwidth. Doing so, the data spectrum becomes folded in frequency domain, and likewise,  the interference spectrum is also folded. This explains the single interference artefact's azimuth spectrum in Fig. \ref{ice_sub1}.

 In S-1 imagery, there are often multiple LFM interference received. Since SAR focusing is a linear operation, the overall interference response in image domain is the sum of each individual interference artefact. The image in Fig. \ref{ice_sub2} shows an example of this situation.

\subsection{Results and Analyses with S-1 Data}
\begin{figure*}[htbp]
  \centering
  \hspace{0.3cm}Original  \hspace{4.6cm} PCA \hspace{5cm} RPCA \hspace{6.3cm}\\
  {\includegraphics[width=5.9cm]{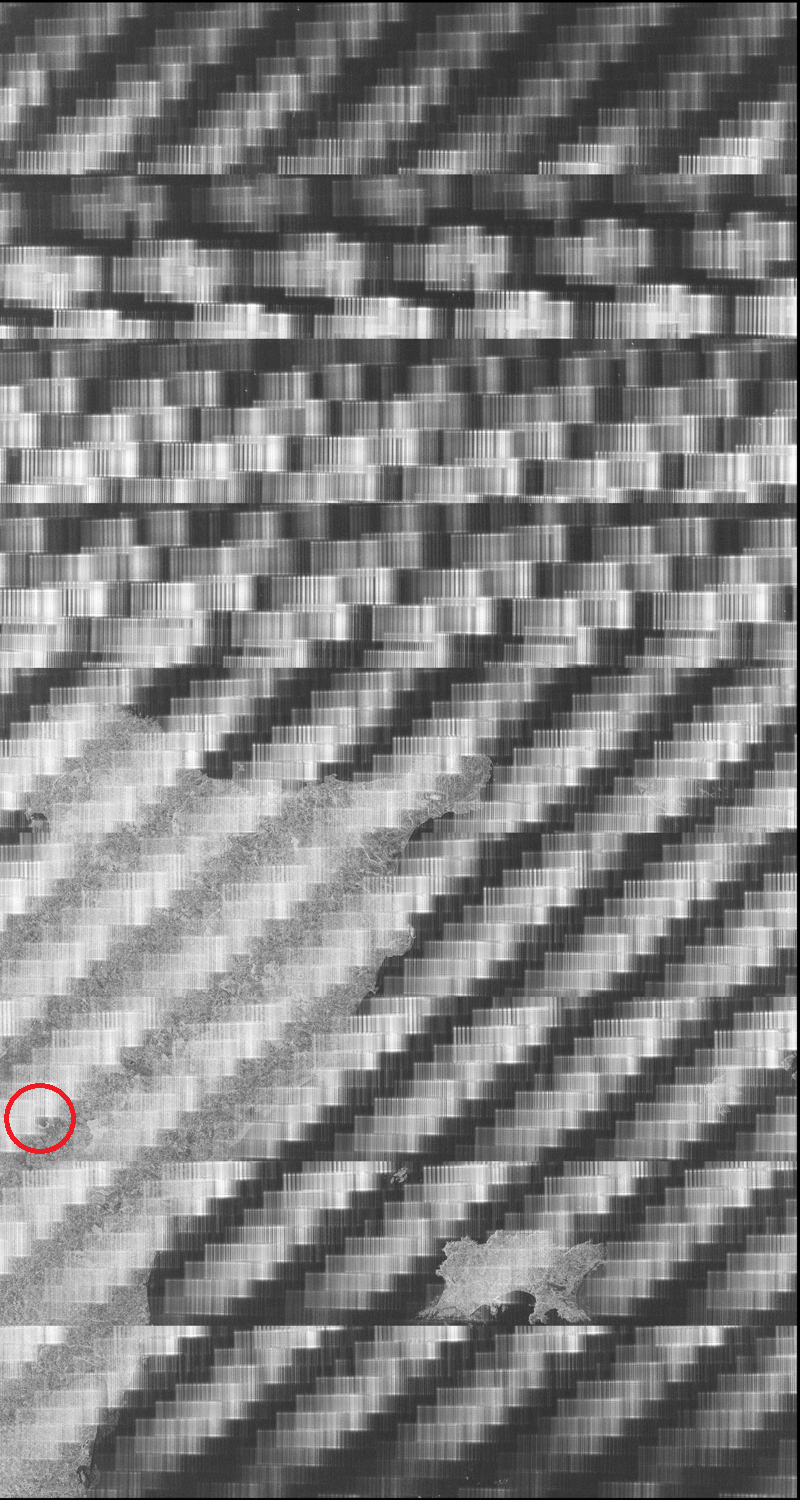}}
  {\includegraphics[width=5.9cm]{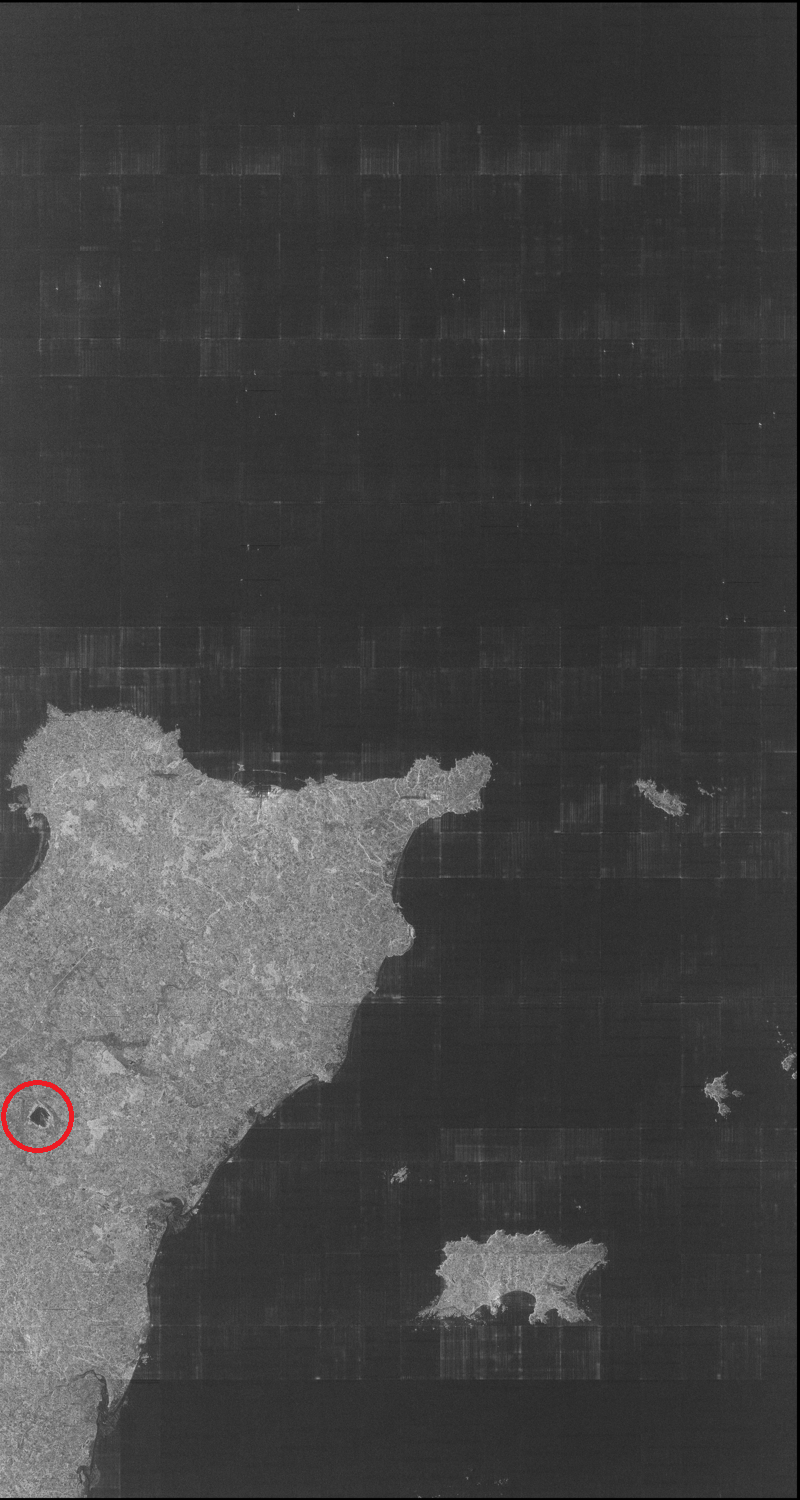}}
  {\includegraphics[width=5.9cm]{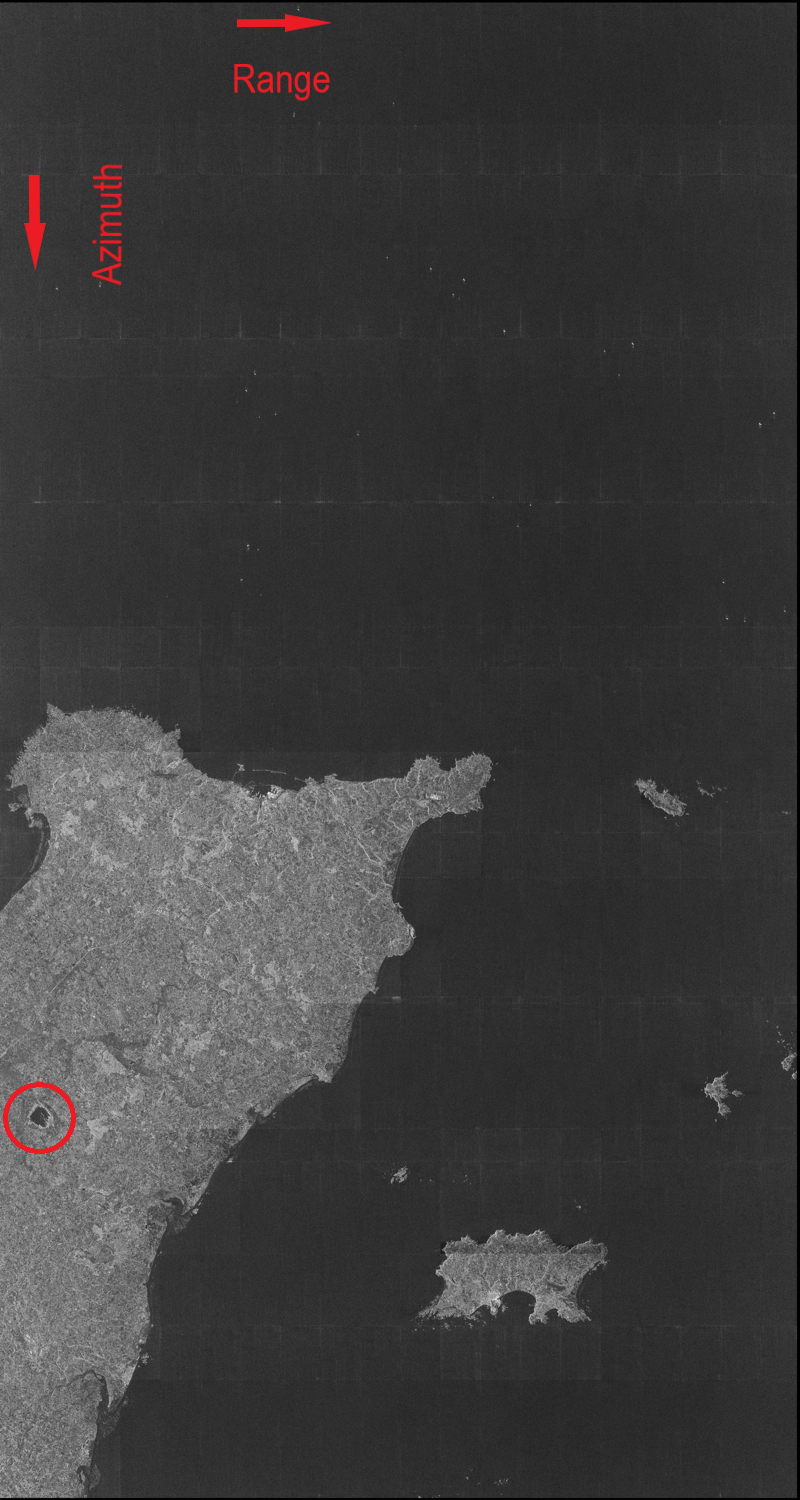}}
  \caption{Processing results on the S-1 IW1 VH image (9 bursts).   Left: original corrupted S-1 IW VH image; Middle: result processed via block-wise PCA ($K=40$); right: result processed via block-wise RPCA. Block size is $1024\times 1024$ pixels. The IW1 image size is $23054 \times12249$ pixels in azimuth and range. For visualization, the three displayed images are resized to $1500\times 800$ pixels. The same colormap scale is used. The total runtime is $68.1$ seconds for PCA and $2.54$ hours for RPCA (with $40$ iterations).}\label{full}
\end{figure*}

\begin{figure*}[htbp]
  \centering
    \subfigure{\begin{sideways}\hspace{1.1cm} PCA, rank 40\end{sideways}\includegraphics[width=17.8cm]{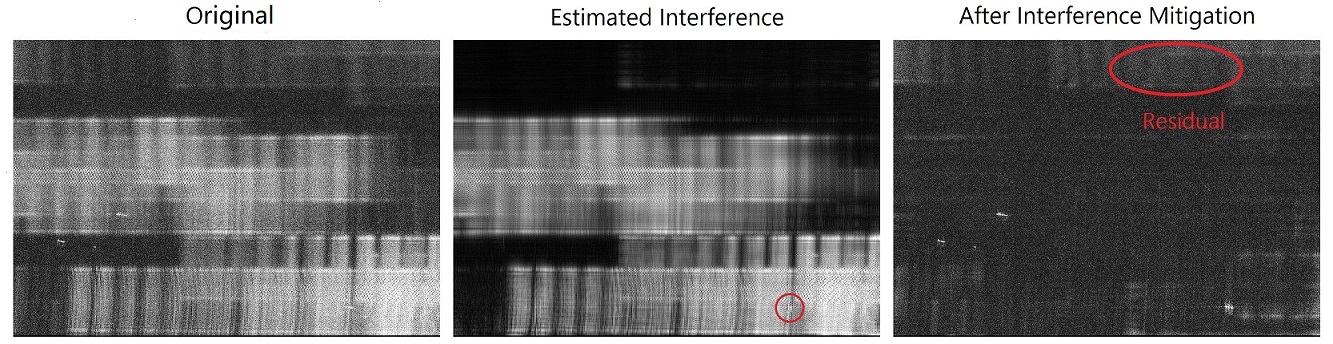}}\vspace{-0.3cm}
    \subfigure{\begin{sideways}\hspace{1.1cm} PCA, rank 80\end{sideways}\includegraphics[width=17.8cm]{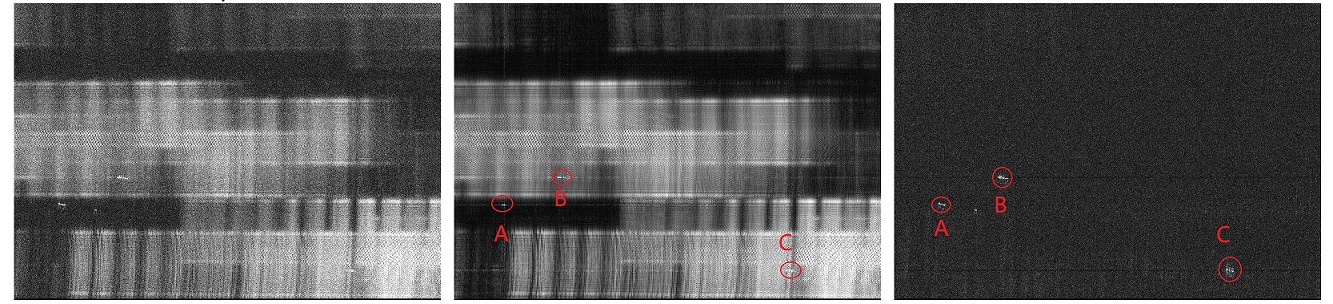}}\vspace{-0.3cm}
    \subfigure{\,\begin{sideways}\hspace{1.6cm} RPCA\end{sideways}\includegraphics[width=17.8cm]{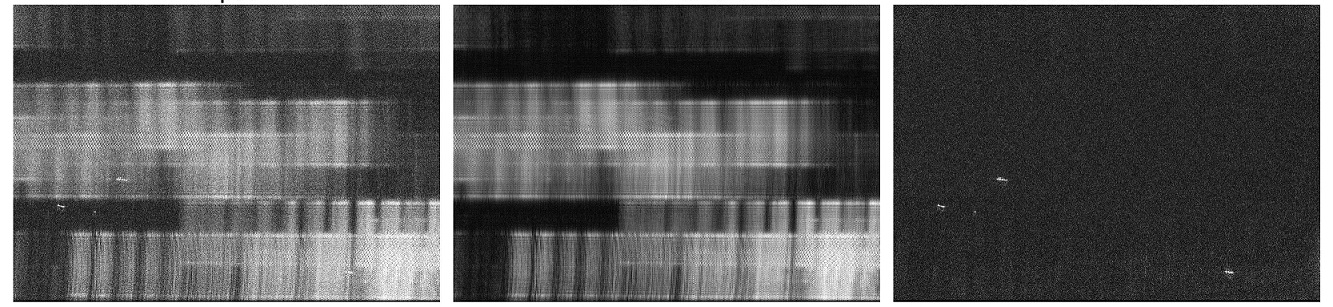}}
  \caption{Processing results on  the selected small patch in the S-1 IW1 image, $1342\times 1935$ pixels. Red circles in the middle column show that parts of ship reflectivity are mistakenly identified by PCA as principal components.}\label{ship}
\end{figure*}
\begin{figure*}[htbp]
  \centering
  \hspace{0.3cm}Original  \hspace{4.6cm} PCA \hspace{5cm} RPCA \hspace{6.3cm}\\
\begin{sideways}\hspace{2cm} VV\hspace{3.5cm} VH\end{sideways}\;\includegraphics[width=17.8cm]{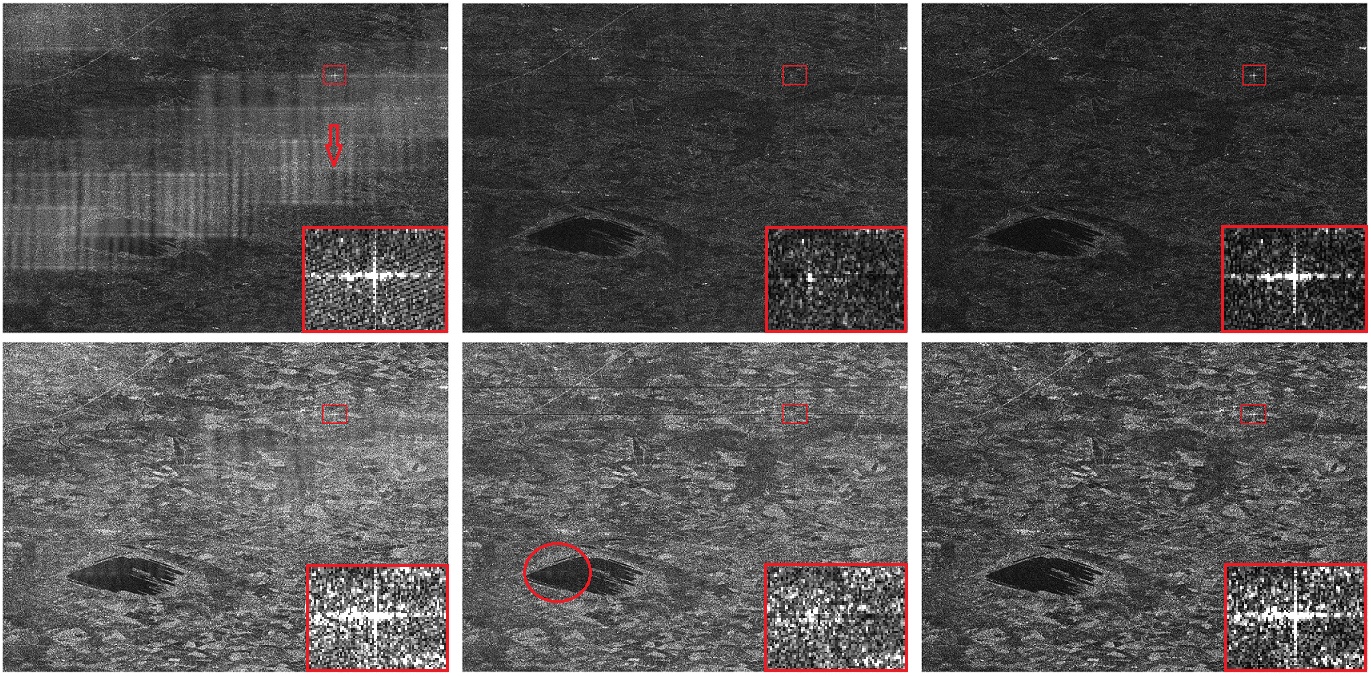}
  \caption{Processing results with S-1 VH and VV data over land. The original VV image has a higher signal-to-interference ratio than the original VH image. A strong point target is mistakenly mitigated by PCA ($K=30$). The same colormap scale is used for both polarizations.}\label{pol}
\end{figure*}
\begin{figure*}[htbp]
  \centering
\includegraphics[width=5.17cm]{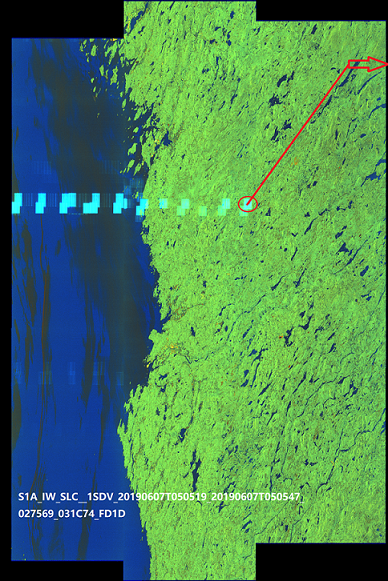}
\includegraphics[width=12.75cm]{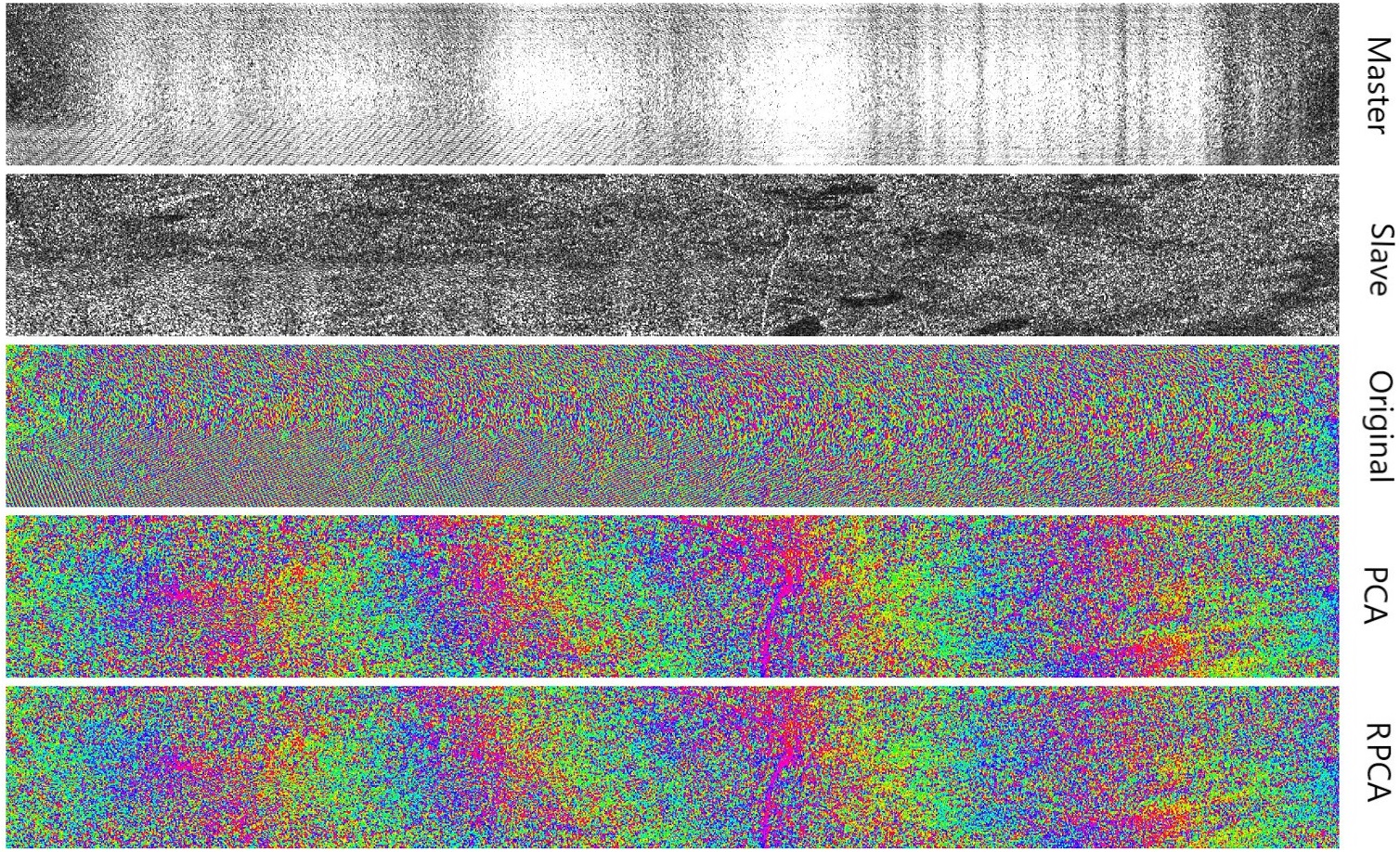}
  \caption{Processing results on S-1 data with SAR interferometry. The left side shows the S-1 quick-look view of the master image. The right side shows (from top to bottom) the master image, the slave image, and the interferometric phases before and after interference mitigation. The topographic and flat-Earth phases are subtracted, and the interferograms are filtered using Goldstein filter with adaptive filter exponent parameter $0.5$, FFT size $64$, and window size $5$. The master image is severely corrupted by interference, and the slave image is not severely corrupted. Each image in the right side has $201\times 1651$ pixels. HSV colormap is used for the visualization of these interferometric phases.}\label{insar}
\end{figure*}

In our experiments, the original S-1 image to be processed is shown by the first image in Fig. \ref{full}. Note that S-1 IW has three sub-swaths, called IW1, IW2 and IW3, respectively, and the image has two polarizations, i.e., VV and VH. The tested data is an IW1 sub-swath image with VH polarization. The data identifier is S1B\_IW\_SLC\_\_1SDV\_20190411T062253\_20190411T06232\\0\_015755\_01D915\_2A6A, and the imaged region in this sub-swath is marked in Fig. \ref{s1_france}. The original size of the IW1 image is  $23054 \times12249$ pixels. For visualization, we reshape it to $1500\times 800$ pixels. In the original image, multiple severe LFM interference artefacts are observed. These artefacts have strong intensity, burying parts of the terrain reflectivity. For example,  the red circled area, which is dominated by interference artefacts, actually contains a lake.

Since the IW1 image has a very large size, it is not practical to apply PCA  and RPCA directly due to the cubic computational complexity of SVD in matrix dimensions, i.e., ${\cal O}((N_a+N_r)\min^2(N_a,N_r))$. To avoid this problem, we adopt block-wise processing. Specifically, we split the whole image into a number of blocks with $1024\times 1024$ pixels, and then use PCA  and RPCA to perform interference mitigation block-by-block. For PCA, we implemented it via $K$-SVD with $K=40$. The processed result is shown by the second image in Fig. \ref{full}, which is also resized to $1500\times 800$ pixels for visualization. In the processed result, we can see that most interference artefacts can be removed well, although there still exist some residual interference. The lake in the red-circled region now is clear to see.
To obtain improved interference mitigation performance, we  process the IW1 image via block-wise RPCA. The processed result is shown by the third image in Fig. \ref{full}. It can be seen that, the interference artefacts are better suppressed compared to the PCA results.

In Fig. \ref{full}, although most interference artefacts are mostly mitigated, there are still limitations for both methods. For example for the island of Jersey, part of the backscattered signal of interest has been removed, creating a radiometric step along the azimuth direction. The reason lies on the partial correlation between the signal of interest in the matrix $\bf I$ and the estimated low-rank interference subspace, leading to a reduction in image amplitude after interference mitigation. These limitations show that the interference can be over-compensated.

To better compare the performances of above two  methods, we select a small region to perform experiments. The selected region contains multiple interference artefacts, under which there are several ships. The results processed by PCA with $K=40$ and $80$ are shown in the first two rows in Fig. \ref{ship}, respectively, and the results processed by RPCA are shown in the third row. In the first row, we find that removing top $40$ principal components, which is taken as interference contributions, still produces significant residual interference, as marked by the red circle. So, a straightforward way to further mitigate the interference artefacts is to use a larger $K$, as the result in the second row shows (with $K=80$). However, it can be seen that parts of the ship reflectivity is also identified by PCA as principal components in this setting, as shown by the red circles marked with A, B and C. This behavior will result in an estimation error for the reflectivity profile of the ships. By RPCA, the result has a similar residual level to the PCA result in the second row, but the ship reflectivity is not identified as interference components anymore. Thus, RPCA better preserves these strong point scatterers. This advantage is beneficial to point scatterer-based applications like persistent scatterer interferometry, ship detection and classification.

In addition to the above example of ocean region, Fig. \ref{pol} shows some processing results over land for both VV and VH polarizations. In the original data, the interference artefacts in the VV image appear less severe than those in the VH image due to stronger backscatter in this polarisation, i.e., the VV and VH images have different signal-to-interference ratio (SIR). By applying PCA and RPCA, the interference artefacts are predominantly mitigated, but there are some differences concerning point scatterers and residual interference. As shown in the figure, the signal from a strong point-like target is partially removed with PCA in both VV and VH images. In addition, there are residual interference in the VV images, as marked by the red circle. PCA requires high interference power to estimate and then mitigate the interference artefact, so high SIR can reduce the mitigation performance. This explains why the VV image has more residual than the VH images in the central column of Fig.16. With RPCA, the interference signals are better cancelled and the point scatterer is well preserved, but the signal of interest is over-compensated, causing a radiometric bias.

In the above experiments, only image amplitudes are presented, and whether the methods preserve phase is not investigated. Referring to (\ref{pca}) and (\ref{rpca}), PCA and RPCA are used to reconstruct the interference-free SLC image, so the two methods are expected to preserve phase information.  To investigate this point, we show another example in Fig. \ref{insar}. In this example, the master image is severely corrupted by LFM interference, and the slave image is not severely corrupted. The associated data identifiers are S1A\_IW\_SLC\_\_1SDV\_20190607T050519\_20190607T05054\\7\_027569\_031C74\_FD1D and S1B\_IW\_SLC\_\_1SDV\_20190\\601T050438\_20190601T050505\_016498\_01F0D5\_F91C, respectively. We mitigate the interference artefacts in the master image and then perform SAR interferometry with the slave image to provide a preliminary assessment on the capability to preserve the phase and use the datasets for InSAR. The interferometric phase before interference mitigation is shown in the third image on the right-hand side of Fig.\ref{insar}. Due to the presence of strong interference, the original interferometric fringes are hardly visible. The interferogram images after interference mitigation with PCA and RPCA are shown on the fourth and the fifth row, respectively. As we can see, the interferometric fringes are now clearly visible, demonstrating that the RFI mitigation techniques proposed in this paper could potentially be exploited to include images contaminated with LFM interference in a SAR interferometric analysis.

It is noteworthy that the two processing methods have much different computational cost. In our implementation, the total runtime for processing the IW1 VH image in Fig. \ref{full} is $68.1$ seconds via PCA and $2.54$ hours via RPCA (on Intel i7-7700 3.6 GHz CPU and 16 GB RAM). The heavy computational cost of the latter makes it unsuitable for processing images that have large corrupted areas. Considering this point, we recommend PCA as the priority.

\section{Conclusion}

In this paper, we investigated the mutual interference problem of two spaceborne SARs where the transmitting LFM signal of a SAR is directly received by another SAR. Based on wavenumber domain analysis, we demonstrated that the image-domain radiometric artefact of a single LFM interference is approximately a two-dimensional LFM signal. It has a limited range and spatial extent, as our theoretical model predicts and verified by simulations in a C-band setting. Based on range-azimuth decoupling approximation and two-dimensional high-order Taylor expansion, we showed that the radiometric interference artefact is low-rank, so that PCA and RPCA can be used for interference mitigation in image domain.  PCA has the advantage of fast processing speed, which costs about $68$ seconds for processing one sub-swath of S-1 IW SLC image via block-wise processing, and thus, we recommend it for practical use. Finally, experiments on S-1 data show that severe radiometric artefacts caused by LFM mutual interference in S-1 SLC images can be significantly mitigated using the proposed methods.
Meanwhile, we should point out that the proposed methods still have some limitations, i.e., producing over-compensated amplitudes and mitigating strong point targets.

It is worth mentioning that  surface sources can also cause LFM interference to spaceborne SARs. Examples of these sources are ship or military air defense radars, which unfortunately sometimes operate at C-band frequencies normally reserved for space SAR missions. Fortunately, the proposed methods can also be used in these situations as long as the interference is an LFM signal.

In spaceborne SAR imaging, another type of mutual interference is the terrain scattered interference (TSI), which occurs when an area is simultaneously illuminated by different SAR systems\cite{s1mpc,s1mpc2,Mingliang2019Mitigation}. The radiometric artefact caused by TSI is significantly different from that caused by LFM mutual interference investigated in this paper due to modulation by the illuminated area. In the future, we would like to investigate mitigation methods for TSI.

\section*{Acknowledgement}

The authors would like to thank the anonymous reviewers for their insightful and constructive comments, and thank European Space Agency for providing open S-1 SAR data.

\bibliographystyle{IEEEtran}
\bibliography{refs}

\end{document}